\documentclass{article}
\usepackage{graphicx} 
\usepackage{amssymb}
\usepackage{amsmath}
\usepackage{amsthm}
\usepackage{color}
\usepackage{comment}
\usepackage{enumitem}		
\usepackage{geometry}		
\usepackage{graphicx}
\usepackage{authblk}
\usepackage{amssymb}
\usepackage{listings}
\usepackage{xcolor}
\usepackage{caption}
\usepackage[linesnumbered,ruled,vlined]{algorithm2e}
\usepackage{color}
\usepackage{algpseudocode}
\usepackage{subcaption}
\usepackage{hyperref}
\usepackage{url}
\newtheorem{theorem}{Theorem}
\title{Complexity of coexistence regions in the GRHT map}
\author[a]{Sishu Shankar Muni}
\affil[a]{School of Digital Sciences, Digital University Kerala, Technopark phase-IV Campus, Mangalapuram, 695317, Thiruvananthapuram, India}
\usepackage[english]{babel}

\begin{document}

\maketitle
\begin{abstract}
GRHT map refers to a planar map which showcases the coexistence of infinitely many stable periodic orbits via the phenomenon of Globally Resonant Homoclinic Tangencies. This paper investigates the geometric properties of coexistence regions in the case of codimension-three scenario. We introduce parameters into the non-invertible GRHT map to understand the unfolding behavior of the GRHT. Near the infinite coexistence regions, there exists series of codimension-one saddle-node and period-doubling bifurcations. The most common overlapping region in the parameter space reveals the parameters with which there can be coexisting periodic orbits. Various slices of parameter space are considered to understand the coexistence regions in two-dimensional parameter space.   We show that the parameter region of coexistence are polygons and are convex sets. We develop an algorithm that detects the number of vertices of the most common overlapping regions via optimization techniques.  We study the variation of the number of vertices and area of the most common overlapping region with the variation in parameters of the map. It illustrates that when the number of coexisting stable periodic orbits increase, the area of the most common overlapping region decreases. Moreover the number of vertices of the most common overlapping region increases as the number of coexisting stable periodic orbits increases. We also explore the variation of the area and the number of vertices of the most common overlapping region with the simultaneous variation of two parameters of the GRHT map. It reveals that the variation in the number of vertices of the most common overlapping region is not trivial and varies in a highly nonlinear fashion illustrating the geometric complexity of the coexisting regions of stable periodic orbits in the parameter space.
\end{abstract}

\section*{Introduction}
Multistability refers to the coexistence of attractors for a particular set of parameters in a dynamical system \cite{MENDES2000}. Multistability is exhibited by nonlinear dynamical systems.  Multistability can be realized via many techniques depending on the dimension of the dynamical systems considered. In the case of one-dimensional maps, cobwebs can identify the presence of coexisting chaotic attractors, periodic attractors \cite{Cassal-Quiroga2022-kd}. One can run the cobweb plots for different initial conditions  to account for different cobwebs for the same parameter values, accounting for multistability. For planar maps, we can use a pictorial representation of basin of attraction to understand the presence of number of coexisting attractors and which initial condition can lead to a particular coexisting dynamical attractor. Basin of attraction plots are created on the two-dimensional plot of initial conditions. A $1000 \times 1000$ grid of initial conditions are taken on $x$ and $y$ axis. For each instance of initial condition, the system is run over a  long enough iterations and final iterations are considered discarding transients and is checked whether the attractor is periodic, chaotic, or the iterates are diverging. Usually such basin of attraction in the case of nonlinear systems are very complicated and are usually fractals \cite{Nusse1994}. The border of fractal is bounded by the stable manifolds of saddle orbits \cite{McDonald1985}. For three-dimensional map, till date there is no reliable method yet to understand multistability except considering various slices of the basin of attraction and phase portraits for various initial conditions. For higher dimensions, one can consider phase portrait for various initial conditions. Basin of attraction is also useful in detecting multistability in higher dimensional network of oscillators. In a ring-star network of Chua oscillators, coexistence of single-well and double-well chimera states were found and were illustrated via the riddled fractal basin of attraction plots \cite{dosSantos2023}. However, interestingly irrespective of the dimension of the the dynamical system considered, one can always construct a one-parameter and two-parameter bifurcation diagram to understand the coexistence of attractors as they give non-overlapping regions which can hint presence of multistable regions in the parameter space. They have been useful in detecting multistability in neuronal mappings \cite{MuniIZH,MuniHR}, memristive systems \cite{Kengne2022,MuniMem}, biological mappings \cite{VS2025}.  

Coexistence of infinitely many chaotic attractors were discussed in case of a memristive circuit  \cite{Bao2016}. In many of previous literature, coexistence of infinitely many chaotic attractors were found in applied systems such as neuron systems \cite{BouiABoya2024,NjitackeTabekoueng2022}. In \cite{ItaloBischi2000}, Researchers have shown the presence of multistable behavior (existence of many coexisting attractors including cycles and chaotic attractors)  in the Cournot duopoly game and is proven to be a characteristic property of such game theoretic mappings.  The oligopoly model modelled via a three dimensional map also exhibited several coexisting Nash equilibria \cite{NabihAgiza1999}.
Interestingly, there has been very limited work on coexistence of stable periodic orbits.  However, coexistence of more than four to five stable periodic orbits is not common. In the thesis \cite{GRHT}, necessary and sufficient conditions for a planar mapping to exhibit infinitely many stable periodic orbits were discussed. Moreover an explicit mapping was shown which exhibits the coexistence of infinitely many stable periodic orbits. The mechanism behind such coexistence was coined to be \textit{Globally Resonant Homoclinic Tangencies} \cite{Muni2022HT} and therefore the planar mapping is referred to as \textit{GRHT map}. 

In \cite{Huang2008}, coexistence of attractors were studied for two-dimensional neural networks with multilevel activation functions. Authors have showed that the system has $n^2$ isolated equilibrium points when the activation function constitutes of $n$ segments. 
Coexistence of attractors has been shown in a class of neural networks with Mexican-hat type of activation functions \cite{LiliWang2012}. A set of sufficient conditions were also presented for the occurence of multistability. Prediction of multistability was carried via a center manifold analysis for a model of pair of neurons in the presence of time-delayed connections \cite{Shayer2000}. The multistability property of a Cellular  Nonlinear Network (CNN) was used to extract specific regions of a multimedia image \cite{Morfu2007}. In \cite{Cheng2006}, authors have discussed the existence of multiple stable stationary states for Hopfield Neural Networks with and without delay. Such stable stationary states are related to the memory capacity for neural networks. Multistable behavior in large scale models of brain activity were reported by Golos et. al. \cite{Golos2015}. Coexistence of synchronized and desynchronized states in a time-delayed interaction and pinning force on coupled oscillators were studied \cite{Kim1997}. Researchers have shown the bistable property of lactose utilization network of E.Coli \cite{Ozbudak2004}. Multistable coordination dynamics exists at many levels of information processing. It is found in multifunctional neural circuits and large scale neural circuitry in humans. Researchers in \cite{Kelso2012} reviewed various key evidences in these areas and have shown some theoretical arguments.  Experimentally, researchers have observed multistability in the case of neuronal dynamics \cite{Maistrenko2007} in whcih they illustrate the coexistence of fully synchronized state, fully desynchronized state and a variety of cluster states in a wide range of parameter space. It is shown that such coexistence of attractors can occur only for asymmetric STDP.  

Multistability has also been seen in coupled nonlinear oscillators and has been an active topic over the recent years \cite{Postnov1999}.  Nested structure of phase synchronized regions of different attractor families are observed in coupled nonlinear oscillators. In the modulated laser system and Duffing oscillator system, researchers have shown the presence of multistability  and moreover have shown that such coexistence occurs when the attractor volume becomes large enough to meet other solutions \cite{Mehrabbeik2023}.  Multistability has been illustrated in a simple rock-paper-scissor competition model \cite{Park2018}. They also show that the mechanism of the occurence of  multistability is associated via the occurence of subcritical Hopf bifurcation. Moreover, this important and unexpected finding will open up an opportunity to interpret rich dynamical phenomena in ecosystems  via various types of interactions and competitions among ecological species. 

Slow periodic modulations with adjusted amplitudes and frequencies  \cite{GOSWAMI2008} can create significant qualitative changes in the dynamics and also control the number of coexisting attractors. Five coexisting chaotic attractors were illustrated in the forced Duffing equation\cite{Arecchi1985}. Such multistability is referred to as generalized multistability. A delayed feedback control mechanism has been useful to account for multistability and has been applied to the integrate and Fire neuron model   \cite{Foss1997}. Coexisting oscillatory and silent regimes are exhibited by neuronal and cardiac systems. Such coexistence has been related to play an important role in short term memroy and posture. It is advantageous in the case of multifunctional central pattern generators. Six different types of multistabilities were discussed \cite{Malashchenko2011}. It was observed that the human alpha rythm (one of the prominent attributes of cortical activity) bursts erratically between two distinct modes of activity. Furthermore, a detailed mechanism for this multistable phenomenon were characterized \cite{Freyer2011}. Extreme multistability was used in the field of secure communication as these systems are highly unpredictable with respect to initial conditions. They have also shown that they synchronization attacks are ineffective in these cases \cite{Kelso2012}. 

Gavrilov et.al. showcased that there exists a sequence of coexisting periodic orbits near the homoclinic tangencies. However, generically all such coexisting periodic orbits are unstable \cite{Gavrilov1973}. As a motivation to describe a mechanism for multistability originating from coexisitng sequence of stable periodic orbits at a homoclinic tangency, it was found that few additional conditions can create a sequence of stable periodic orbits.  \textit{Globally Resonant Homoclinic Tangencies} are hubs for extreme multistability.  In the presence of homoclinic tangencies with additional necessary and sufficient conditions, the coexistence of infinite sequence of stable periodic orbits are guaranteed and such special tangencies are referred to as \textit{Globally Resonant Homoclinic Tangencies}. A detailed theoretical mechanism was formulated \cite{GRHT}.  

It is further interesting to understand the regions where such infinite coexistence occurs. For the phenomenon to occur, we need minimum codimension-three or three fundamental parameters in the planar map. Thus the region of such infinite coexistence can be only seen in three dimensions or higher. To begin with simple case, we consider various slices of the three dimensional parameter space. Usually there exists a parameter region in two-dimensional parameter space where such infinite coexistence of asymptotically stable periodic orbits can exist. Interestingly, we show in this paper that for the planar map considered, such regions of coexistence are analytically tractable. Computing parameter region and getting analytical expressions for coexistence boundaries are seldom in the field of nonlinear dynamics. Previous works on Globally Resonant Homoclinic Tangencies focused on the illustration of the infinitely many stable periodic orbits along with theoretical necessary and sufficient conditions. However, detailed discussion on the shape and areas of the parameter sets in 2D parameter spaces were absent. In continuation of the previous work, we discuss on the geometry of these regions of coexistence on the two-dimensional parameter space and understand how these shapes vary with the variation of parameters. Moreover, we show how the area of the parameter region with coexistence of stable periodic orbits approach zero as number of coexistence of periodic orbits increases. This highlights the reason why such infinite coexistence of stable periodic orbits are so scarce to observe but are exotic at the same instance. 

The main contributions of the paper are outlined as follows:
\begin{itemize}
\item We show the coexistence region of periodic orbits as polygons in 2D parameter space where the coexistence of stable periodic orbits are present.
\item An analytical algorithm is presented to automate the detection of vertices of the coexisting parameter regions.
\item Sudden variations in the vertices of the polygon are illustrated over the variation of two-parameters which illustrates the complex structure of the region of multistability.
\item Three-dimensional stability region is showcased along with its variation of the number of such coexisting periodic orbits.
\end{itemize}
 
The paper is organised as follows. 
In \S \ref{sec:GRHTmap}, the planar GRHT map is introduced and properties of the map is studied like non-invertibility and also infinite coexistence of stable periodic orbits are illustrated. In \S \ref{sec:Perturbation}, perturbations to the Globally Resonant Homoclinic Tangency (GRHT) is discussed and recaptitulated. Sequences of saddle-node and period-doubling bifurcations are illustrated due to the perturbations. Also computation of those bifurcation points via a modified bisection method is discussed. This is important as these codimension-one bifurcation curves bound the stability regions on the parameter space. In \S \ref{sec:complexity}, complexity of the coexistence regions are discussed. Mainly, we showcase that the coexistence regions are polygons with number of vertices changing drastically with the increase in the number of coexisting stable periodic orbits. Moreover, we discuss about the area of the polygon representing the most common overlapping region and discuss its variaiton with increase in the number of coexisting stable periodic orbits. The paper concludes with future directions and conclusions.

\section{GRHT map and infinite coexistence}
\label{sec:GRHTmap}
Homoclinic tangencies refer to the tangential intersection between the one-dimensional stable and unstable manifolds of an invariant set (simplest case: saddle fixed point). Generically, near such homoclinic tangencies there exists a sequence of coexisting unstable periodic orbits \cite{GaSi72}. In 1974, NewHouse had shown that two-dimensional maps can exhibit infinitely many attractors near homoclinic tangencies \cite{Nhouse}. However, no explicit examples of smooth maps were known till 2020. This motivates one to ask the question: under what conditions can there be infinite co-existence of stable periodic orbits at the homoclinic tangency? The answer to this question is affirmative and was considered by Muni et. al. in 2022 \cite{GRHT}. The authors prove that it is indeed possible for a 2D planar map with a homoclinic tangency to a saddle fixed point to develop coexistence of infinitely many stable periodic orbits. Furthermore, the codimension of this phenomenon depends on the nature of the saddle:
\begin{itemize}
    \item Codimension-four if the saddle is orientation-preserving.
     \item Codimension-three if the saddle is orientation-reversing.
\end{itemize}
These results establish new conditions under which NewHouse-like behavior can occur, but with an infinite number of stable periodic orbits, rather than just strange attractors or sinks. The phenomenon of such infinite coexistence of stable periodic orbits is referred to as Globally Resonant Homoclinic Tangencies (GRHT). We recall a concrete instance of the 2D map which showcases  \textit{Globally Resonant Homoclinic Tangencies} that develops infinite coexistence of stable periodic orbits. The map is constructed in such a way that it is linear in a neighborhood of the origin and the saddle fixed point has a homoclinic tangency (tangential intersection of the 1D stable and unstable manifolds emanating from the saddle fixed point along the stable and unstable eigendirections). 

\begin{figure}[!htbp]
\begin{center}
\includegraphics{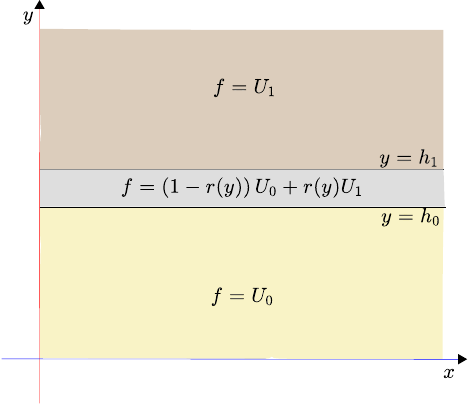}
\caption[Phase space of map exhibiting GRHT phenomenon.]{  Phase space structure of the piecewise-defined map \eqref{eq:fEx}. The space is divided into three horizontal strips based on the value of $y$. For $y < h_0$, the map is defined as $f = U_0$ (yellow region), and for $y > h_1$, it is $f = U_1$ (brown region). In the intermediate zone $h_0 < y < h_1$, the map smoothly interpolates using a convex combination $f = (1 - r(y)) U_0 + r(y) U_1$ (gray region). The lines $y = h_0$ and $y = h_1$ demarcate these regions. Colors help visually separate the function domains.}
\label{fig:smooth_example}
\end{center}
\end{figure}

The explicit form of the piecewise-smooth $C^{1}$ family of maps is given as follows
\begin{equation}
f(x,y) = U_0(x,y) + w(y) \left[ U_1(x,y) - U_0(x,y) \right]
\label{eq:fEx}
\end{equation}
where  weighting function $w(y)$ is given by
\[
w(y) =
\begin{cases}
0, & y \leq h_0 \\
r(y), & h_0 \leq y \leq h_1 \\
1, & y \geq h_1
\end{cases}
\]  
where the expressions of $U_{0}, U_{1}, r(y)$ are shown in \eqref{eq:U0}, \eqref{eq:U1}, \eqref{eq:r} respectively, and
\begin{align}
h_0 &= \frac{2 |\lambda| + 1}{3}, &
h_1 &= \frac{|\lambda| + 2}{3},
\label{eq:xi0xi1}
\end{align}
with $|\lambda|<1$. This formulation above shows smooth transition of the function $f(x,y)$ from $U_{0}$ to $U_{1}$.
 The piecewise effect of \eqref{eq:fEx} is shown in Fig.~\ref{fig:smooth_example}.

We let
\begin{equation}
U_0(x,y) = \begin{bmatrix}
\lambda x \\
\sigma y
\end{bmatrix},
\label{eq:U0}
\end{equation}
 and
\begin{equation}
U_1(x,y) = \begin{bmatrix}
1 + c_2 (y - 1) \\
d_1 x + d_5 (y - 1)^2
\end{bmatrix},
\label{eq:U1}
\end{equation}
Considering the simplest case, we define \( r \) such that \( f \) remains \( C^1 \). There is no restriction in choosing $f$ to be $C^\infty$. The latter case just can introduce complicated function $r$ which can limit the analytical tractability of the probelm. For $f$ to be $C^1$, this imposes the following conditions:  
\[
r(h_0) = 0, \quad r'(h_0) = 0, \quad r(h_1) = 1, \quad \text{and} \quad r'(h_1) = 0.
\]  
The unique polynomial \( r \) which satisfy the above conditions are given by  
\begin{equation}
r(y) = s \left( \frac{y - h_0}{h_1 - h_0} \right),
\label{eq:r}
\end{equation}  
where  
\begin{equation}
s(z) = 3z^2 - 2z^3.
\label{eq:s}
\end{equation}

\subsection{Non-invertibility}
\begin{figure}[t!]
\begin{center}
\includegraphics[width=0.95\textwidth]{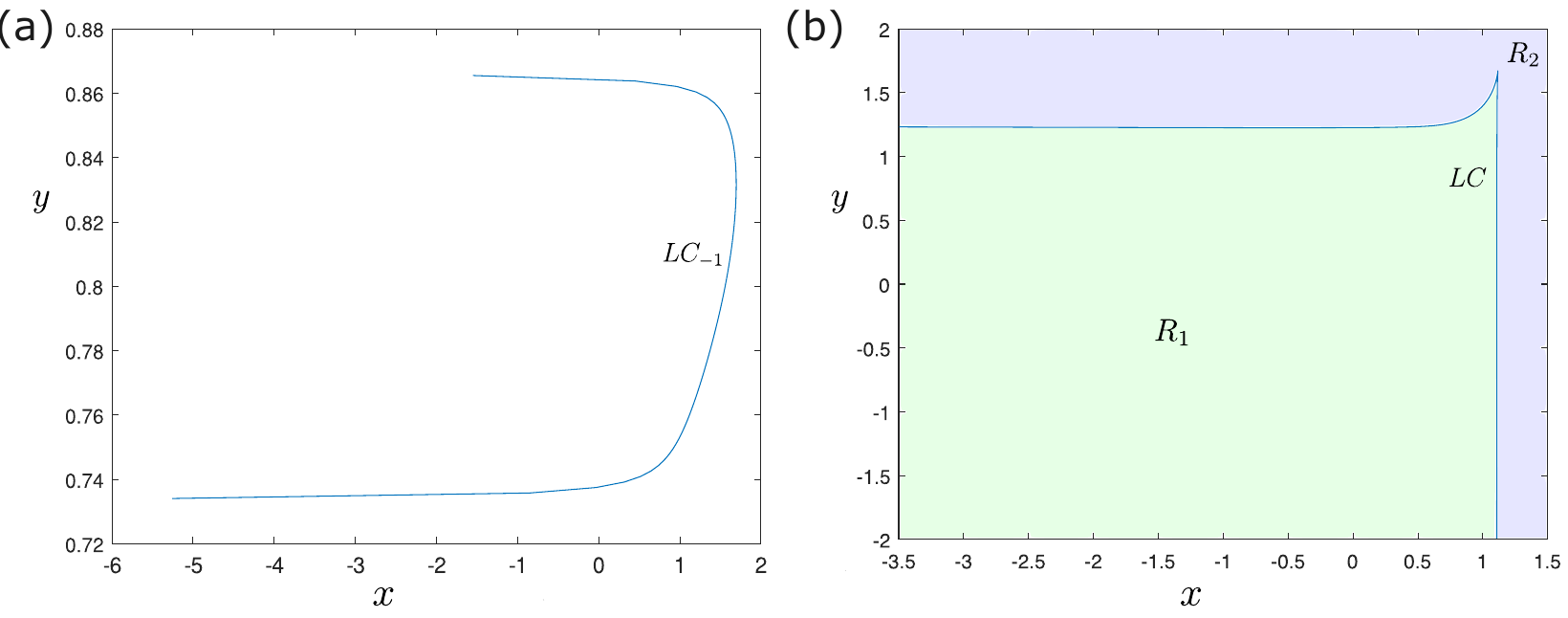}
\caption[Critical curve of the toy map.]{ \textcolor{black}{The critical curve \( LC \) of the map \( f \). In (a), $LC_{-1}$ curve is shown. In (b), the critical curve $LC$ is shown which separates region into $R_1$ consisting of $3$ preimages and region $R_2$ consisting of $1$ preimage. The non-invertible map \( f \) belongs to the \( Z_1 - Z_3 \) type. The parameters are chosen as \( \lambda = \frac{3}{5}, \sigma = \frac{5}{3}, d_1 = 1, c_2 = -0.8, d_5 = 0.8 \).}}
\label{fig:LCToymap}
\end{center}
\end{figure}

Non-invertibility of a dynamical system is responsible for the stretching and folding property \cite{CM96}. It is important to notice that the GRHT map \eqref{eq:fEx} under consideration is non-invertible. Observe that  under the discrete mapping $f$, there exist two distinct points $(0,1)$ and $(\frac{1}{\lambda},0)$ which maps to $(1,0)$. We set $\rm{det}(f) = 0$, see Fig. \ref{fig:LCToymap} (a) to obtain the critical curve which separates the phase space regions according to the distinct number of preimages  shown in Fig. \ref{fig:LCToymap}. Observe that the critical curve forms a cusp located near $(x,y) \approx (1.6706,1.1156)$. This curve separates the distinct number of preimages on the plane, see Fig. \ref{fig:LCToymap} (b). We next compute the number of preimages on either side of the curve $LC$. The region $R_{1}$ has three preimages whereas the region $R_{2}$ has one preimage. Therefore, we can conclude that the GRHT map \eqref{eq:fEx} is of type $Z_{1}-Z_{3}$.

\subsection{Infinite coexistence}
 The fixed point of \( U_0 \) is located at \( (0,0) \), and it exhibits saddle-like behavior. The fixed points of the transformation \( U_1 \) can be determined through solving the following set of equations:  

\[
    \begin{aligned}
    1 + c_{2}(y-1) &= x,\\
    d_{1} x + d_{5}(y-1)^2 - 1 &= y-1,
    \end{aligned}
\]
\(\label{eq:U1fpseqn}\)

Solving these equations yield two fixed points:  

\[
    \begin{aligned}
    y^{*} &= 1 + \frac{(1-c_{2}d_{1} \pm \sqrt{(c_{2}d_{1})^2 - 4d_{5}(d_{1}-1)})}{2d_{5}},\\
    x^{*} &= 1 + c_{2}(y^* - 1).
    \end{aligned}
\]
\(\label{eq:U1fps}\)

\begin{figure}[h!]
\begin{center}
\includegraphics[width=0.75\textwidth]{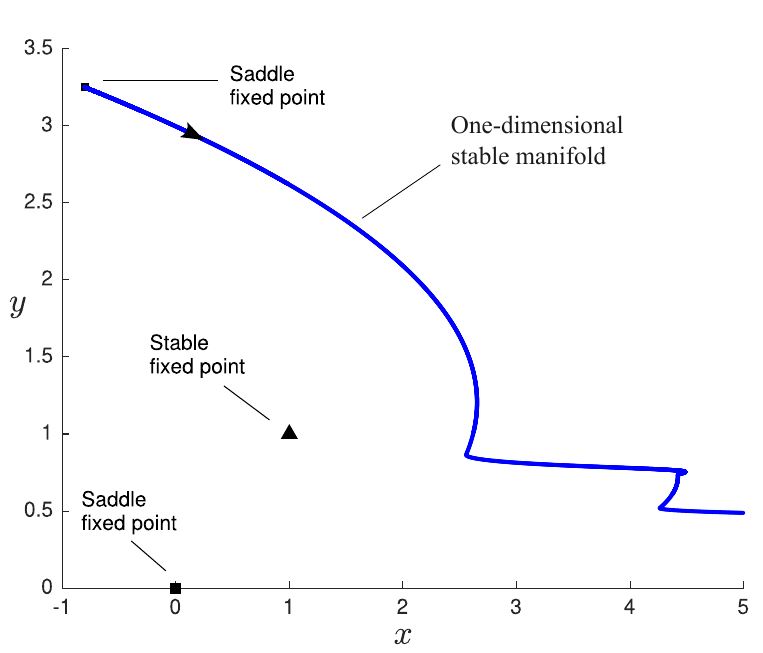}
\caption[Stable manifold of a saddle fixed point of map $U_{1}$. ]{The stable manifold in blue of the saddle fixed point $(-0.8,3.25)$ of map $U_{1}$. The stable fixed point $(1,1)$ of map $U_{1}$ is marked in black. The saddle fixed point $(0,0)$ of the map $U_{0}$ is also shown. The parameters are set as $\lambda = \frac{3}{5}, \sigma = \frac{5}{3}, d_{1}=1, c_{2}=-0.8, d_{5} = 0.8$.}
\label{fig:StableManifoldU1}
\end{center}
\end{figure}

For the given parameter values \( d_{5} = 0.8 \), \( c_{2} = -0.8 \), \( \lambda = \frac{3}{5} \), \( \sigma = \frac{5}{3} \), and \( d_{1} = 1 \), the fixed points are identified at \( (1,1) \) and \( (-0.8,3.25) \). Analyzing the Jacobian matrix and computing the eigenvalues of \( U_1 \) at these points reveal that \( (1,1) \) is asymptotically stable, whereas \( (-0.8,3.25) \) exhibits saddle point behavior. The corresponding stable manifold of this saddle is illustrated in Fig.~\ref{fig:StableManifoldU1}.

To numerically compute the basins of attraction, we evaluate a \( 1000 \times 1000 \) grid of \( x \) and \( y \) values. Each point within this grid undergoes 1000 iterations under the mapping \( f \), generating an orbit \( (x_i, y_i) \) for \( i = 0,1,\dots,1000 \). The classification of these points is performed by assessing the norm of the difference between \( (x_{1000}, y_{1000}) \) and \( (x_{1000-p}, y_{1000-p}) \) for progressively increasing values of \( p = 1,2,\dots p_{max}\), up to the maximum considered period. If this norm is below a predefined tolerance (set at \( 10^{-13} \)), the initial point \( (x_0, y_0) \) is identified as part of the basin of attraction of a periodic orbit with period \( p \). Points where this norm exceeds a threshold (fixed at \( 10^2 \)) are categorized as diverging. Any point that neither belongs to a periodic orbit nor diverges is assigned to a separate classification.  

\begin{figure}[!htbp]
\begin{center}
\includegraphics[width=0.8\textwidth]{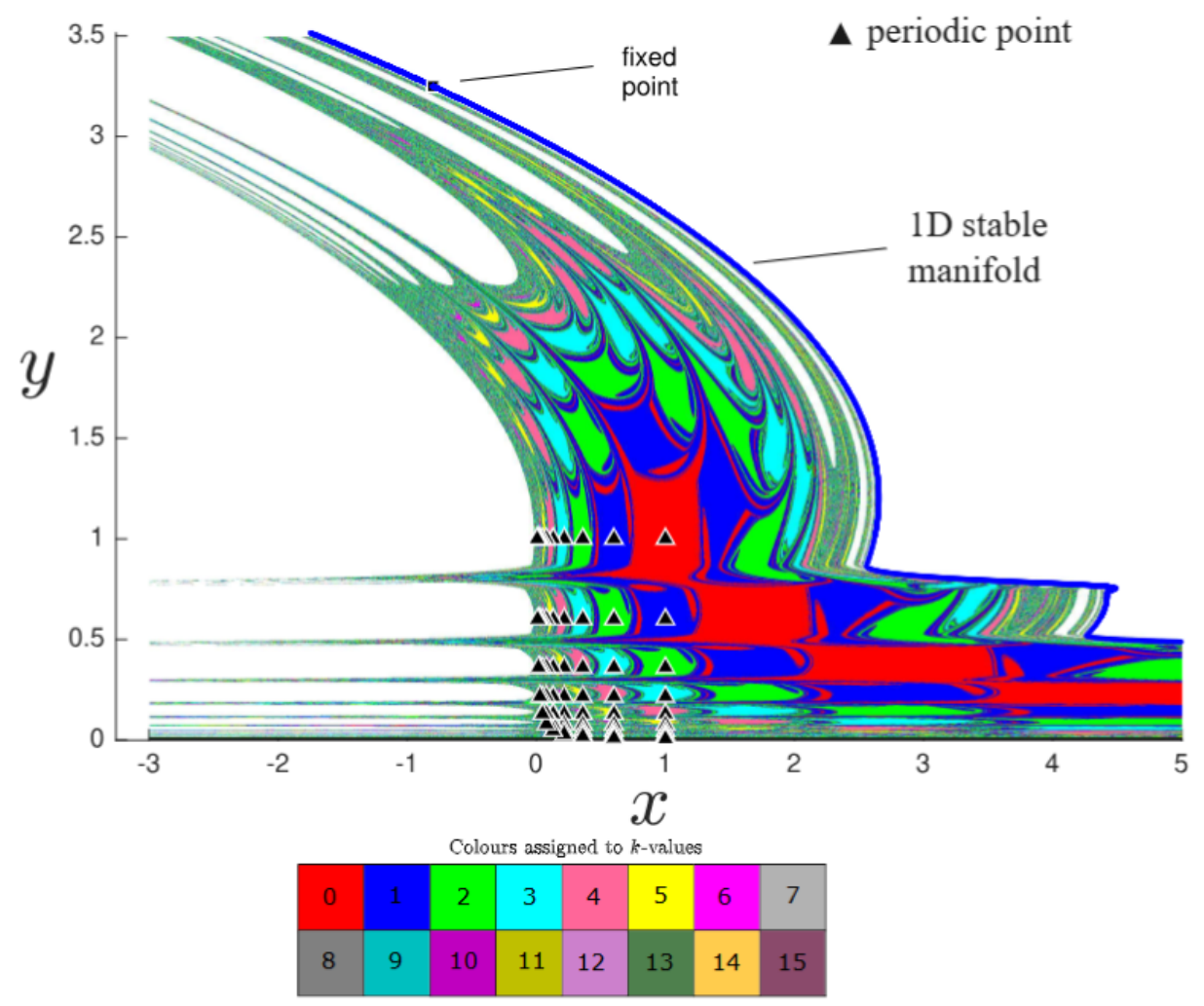}\\
\caption[The stable manifold bounding the basins of attraction]{The basin of attraction for single-round periodic solutions from period 1 to 16 is depicted, with each period distinguished by the colors indicated in the legend. The stable manifold of the saddle fixed point of the map \( U_1 \) seems to form a boundary around the combined basins of attraction. The parameters used are \( \lambda = \frac{3}{5} \), \( \sigma = \frac{5}{3} \), \( d_1 = 1 \), \( c_2 = -0.8 \), and \( d_5 = 0.8 \).}
\label{fig:BasinManifoldU1}
\end{center}
\end{figure}

Figure \ref{fig:BasinManifoldU1} presents the basins of attraction corresponding to single-round periodic solutions for \( k=0 \) to \( k=15 \) (marked with black triangles for different periodic points). \textcolor{black}{It exhibits pronounced multistability, with coexisting single-round periodic attractors of periods $1$ to $16$ (for simplicity, however, one can consider higher periodic orbits as well), each marked in distinct colors as shown in the legend at the bottom. The intricate, interwoven nature of these basins indicates the presence of fractal boundaries, signifying extreme sensitivity to initial conditions. The lower half of the diagram displays stratified horizontal layers, suggesting invariant foliations in the $y$-direction, likely induced by the repeated folding structure of $U_0^k$. The clustering of trajectories near specific narrow bands implies the presence of noninvertible trapping regions, and the coexistence of numerous attractors indicates the occurrence of successive bifurcations.} The findings reveal that the stable manifold of the saddle fixed point at \( (-0.8,3.25) \) acts as a boundary separating the basins of attraction of all single-round periodic solutions. Further extension of this stable manifold suggests that it encompasses the entire white region, particularly within the left half-plane where \( x < 0 \).

\section{Perturbations to Globally Resonant Homoclinic Tangency}
\label{sec:Perturbation}  
We introduce parameters to the example given in \S \ref{sec:GRHTmap}. Specifically, we consider \eqref{eq:fEx} with  

\begin{align}  
U_0(x,y) &= \begin{bmatrix}  
(\lambda + \mu_2) x \big( 1 + (a_1 + \mu_4) x y \big) \\  
\frac{1}{\lambda} y (1 - a_1 x y)  
\end{bmatrix},\\  
U_1(x,y) &= \begin{bmatrix}  
1 + c_2 (y-1) \\  
\mu_1 + (1 + \mu_3) x + d_{5} (y-1)^2  
\end{bmatrix},\\  
h_0 &= \frac{2 |\lambda| + 1}{3},\\  
h_1 &= \frac{|\lambda| + 2}{3}.  
\end{align}  

Fixing the parameters, we set  
\begin{equation}  
\begin{split}  
\lambda &= 0.8, \\  
a_1 &= 0.2, \\  
c_{2} &= -0.5, \\  
d_{5} &= 1,  
\end{split}  
\label{eq:toyParameters}  
\end{equation}  
and vary \( \mu = (\mu_1,\mu_2,\mu_3,\mu_4) \in \mathbb{R}^4 \).  

For \( \mu = O \), equation \eqref{eq:fEx} satisfies the sufficient conditions for infinite co-existence. Since \( \Delta_0 = 2.25 \), it follows that \( \Delta_0 > 0 \) and that the inequality \( -1 < c_{2} < 1 - \frac{\Delta_0}{2} \) holds. Consequently, equation \eqref{eq:fEx} ensures the presence of an asymptotically stable single-round periodic solution for sufficiently large values of \( k \).  

Moreover, such solutions exist for all \( k \geq 1 \), as illustrated in Fig.~\ref{fig:BasinManifoldU1}. Additionally, there is an asymptotically stable fixed point at \( (x,y) = (1,1) \), corresponding to \( k=0 \).  

\subsection{Bifurcation of periodic solutions}  
\label{sec:BifPeriodic}  
A two-parameter bifurcation diagram is constructed to illustrate stability regions, demonstrating that the stable region is enclosed by saddle-node and period-doubling bifurcations. Using the parameter set given in \eqref{eq:toyParameters}, we fix \( \mu_3 = \mu_4 = 0 \) and omit resonance terms by setting \( a_1= b_1 = 0 \). These terms have been excluded in the computations for Fig. \ref{fig:Sevensided} to streamline analytical calculations and facilitate the determination of stable and unstable manifolds, single-round periodic solutions, and bifurcation curves for saddle-node and period-doubling bifurcations. 

\begin{figure}[t!]
\begin{center}
\includegraphics[width=1\textwidth]{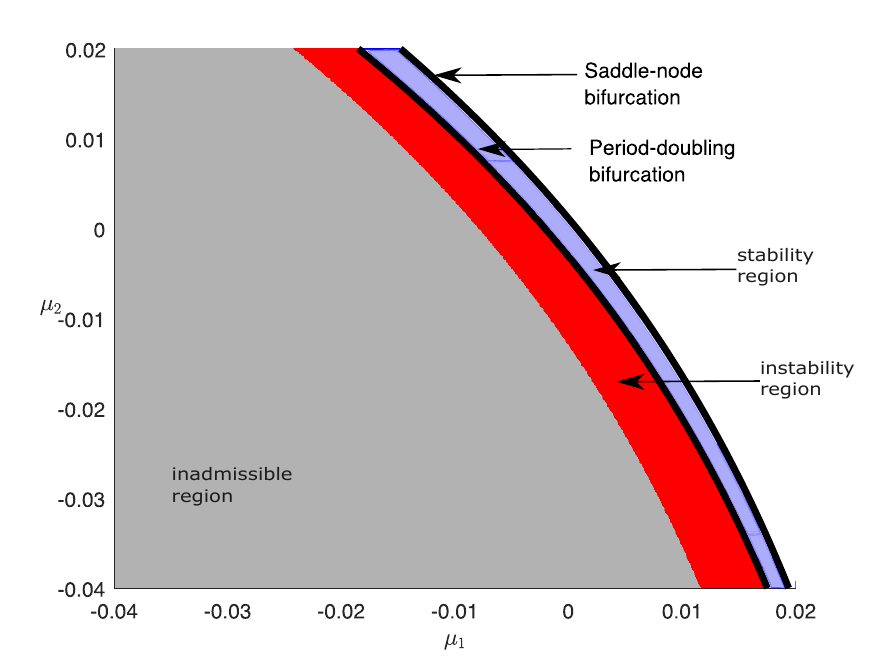}
\caption[Stability region of period-$16$ solution in the $\mu_{1}-\mu_{2}$ parameter plane.]{The stability region of a period-\((k+1)\) solution for \( k=9 \) is analyzed in the \( \mu_1 \)-\( \mu_2 \) parameter plane. The stable region, depicted in blue, is enclosed by the saddle-node and period-doubling bifurcation curves, while the unstable region is shown in red. In the grey area, the periodic solution contains at least one point within the horizontal strip corresponding to the middle segment of \eqref{eq:fEx}. The parameters are set as \( a_{1,0} = b_{1,0} = 0 \), \( \alpha = 0.8 \), \( c_{2,0} = -0.5 \), \( d_{5,0} = 1 \), \( \mu_3 = 0 \), and \( \mu_4 = 0 \).}
\label{fig:mu1mu2SinglePeriodic}
\end{center}
\end{figure}

Fig.~\ref{fig:mu1mu2SinglePeriodic} illustrates the stability region for the map in \eqref{eq:fEx}. The blue region represents the stable zone, while the red region indicates instability in the periodic solution. The grey region corresponds to cases where at least one point of the periodic solution lies within the horizontal strip associated with the middle section of \eqref{eq:fEx}, implying that the periodic solution is not a fixed point of \( U_1 \circ U_0^k \).  \textcolor{black}{The region enclosed between the saddle-node bifurcation curve (outer black line) and the period-doubling bifurcation curve (inner black line) forms a narrow stability band, where a period-10 solution is attracting. This indicates that within this band, eigenvalues of the Jacobian lie strictly within the unit circle. Crossing the inner bifurcation curve (from blue to red) results in a period-doubling bifurcation, marking the onset of instability and the potential emergence of a period-20 orbit. The grey region corresponds to solutions that enter a forbidden domain of the map, specifically those that pass through the “middle strip” of the piecewise system, violating admissibility conditions. Thus, trajectories from this region do not correspond to physically or dynamically valid periodic orbits. The narrowness of the stability region in both $\mu_1$ and $\mu_2$  suggests that precise parameter tuning is required to achieve stable periodic behavior, reinforcing the sensitivity and high-dimensional bifurcation structure of the system. The 2D bifurcation diagram showcases a precise view of how stability for a high-period solution (here, period-10) is gained and lost in parameter space, with clearly demarcated transitions through saddle-node and period-doubling bifurcations. It visually illustrates the system’s multiscale bifurcation structure and  stability regions.}

Given the simplicity of the selected map and parameters, these bifurcations can be determined analytically. \textcolor{black}{For the codimension-three case with $\mu_4=0$, the saddle-node and period-doubling bifurcation curve of $U_{0}^k \circ U_{1}$ has been analytically computed in \cite{Muni2021HT}. We revisit the expressions here which help us to explore the overlapping stability regions of various periodic orbits with variation of $k$.
} 
\begin{theorem}
The saddle-node and period-doubling bifurcation curves of $U_{0}^k \circ U_{1}$ (neglecting resonance terms) are given by \\
(i) Saddle-node bifurcation:

$$
\mu_{1,SN}^{k} = -\frac{(a - c_2 b (\mu_3 + 1))^2 - (a - 1)^2 + 4 d_5 ((\mu_3 + 1)b - 1)c^k}{4 d_5 c^{2k}}.
$$\\
(ii) Period-doubling bifurcation:

$$
\mu_{1,PD}^{k} = -\frac{(a + c_2 b (\mu_3 + 1) - 2)^2 - (a - 1)^2 + 4 d_5 ((\mu_3 + 1)b - 1)c^k}{4 d_5 c^{2k}}.
$$
, where $b = ((\alpha + \mu_2)c)^k$, where $c = \frac{1}{\alpha}$, and $a = c_2 b$. 
\end{theorem}
\begin{proof}
We begin by expressing the composed map $U_0^k \circ U_1$ explicitly:

Let $A = (\alpha + \mu_2)^k$, $C = \left(\frac{1}{\alpha} \right)^k = c^k$, and define $z := y - 1$. Then the fixed point equations are given by:

$$
x = A(1 + c_2 z), \quad y = 1 + z = C \left( \mu_1 + (\mu_3 + 1) x + d_5 z^2 \right).
$$

Substituting $x$ into the second equation yields:

$$
1 + z = C \left[ \mu_1 + (\mu_3 + 1) A (1 + c_2 z) + d_5 z^2 \right].
$$

Expanding and simplifying, we obtain a quadratic in $z$:

\begin{equation}
C d_5 z^2 + C (\mu_3 + 1) A c_2 z - z + C(\mu_1 + (\mu_3 + 1) A) - 1 = 0. \label{eq:1}
\end{equation}

Next, we compute the Jacobian $J$ of the map $U_{0}^{k} \circ U_{1}$ at the fixed point:

$$
J = \begin{bmatrix}
\frac{\partial x'}{\partial x} & \frac{\partial x'}{\partial y} \\
\frac{\partial y'}{\partial x} & \frac{\partial y'}{\partial y}
\end{bmatrix}
= 
\begin{bmatrix}
0 & A c_2 \\
C (\mu_3 + 1) & 2 C d_5 z
\end{bmatrix}.
$$

Hence, the characteristic equation of the Jacobian is:

$$
\lambda^2 - \operatorname{tr}(J) \lambda + \det(J) = 0,
$$

where:

$$
\operatorname{tr}(J) = 2 C d_5 z, \quad \det(J) = -A c_2 C (\mu_3 + 1).
$$

(i) Saddle-node bifurcation: occurs when $\lambda = 1$ is a double root:

$$
\Rightarrow 1^2 - \operatorname{tr}(J) + \det(J) = 0 \Rightarrow \operatorname{tr}(J) = 1 + \det(J).
$$

That is:

\begin{equation}
2 C d_5 z = 1 - A c_2 C (\mu_3 + 1). \label{eq:2}
\end{equation}

Now, plug this value of $z$ into Eq. \eqref{eq:1} to isolate $\mu_1$. From Eq. \eqref{eq:2}:

$$
z = \frac{1 - A c_2 C (\mu_3 + 1)}{2 C d_5}.
$$

Substitute this into Eq. \eqref{eq:1}, and after algebraic simplification (details omitted for brevity), we obtain the saddle-node bifurcation threshold:

\begin{equation}
\mu_{1,SN}^{k} = -\frac{(a - c_2 b (\mu_3 + 1))^2 - (a - 1)^2 + 4 d_5 ((\mu_3 + 1)b - 1)c^k}{4 d_5 c^{2k}}.
\label{eq:SNeqn}
\end{equation}

(ii) Period-doubling bifurcation: occurs when $\lambda = -1$ is a double root:

$$
\Rightarrow 1^2 + \operatorname{tr}(J) + \det(J) = 0 \Rightarrow \operatorname{tr}(J) = -1 - \det(J).
$$

So:

$$
2 C d_5 z = -1 + A c_2 C (\mu_3 + 1).
$$

Hence,

$$
z = \frac{-1 + A c_2 C (\mu_3 + 1)}{2 C d_5}.
$$

Substitute this expression into Eq. \eqref{eq:1}, and again, isolate $\mu_1$. The resulting expression yields the period-doubling bifurcation threshold:

\begin{equation}
\mu_{1,PD}^{k} = -\frac{(a + c_2 b (\mu_3 + 1) - 2)^2 - (a - 1)^2 + 4 d_5 ((\mu_3 + 1)b - 1)c^k}{4 d_5 c^{2k}}.
\label{eq:PDeqn}
\end{equation}
\end{proof}

\begin{figure}[t!]
\begin{center}
\includegraphics[width=1\textwidth]{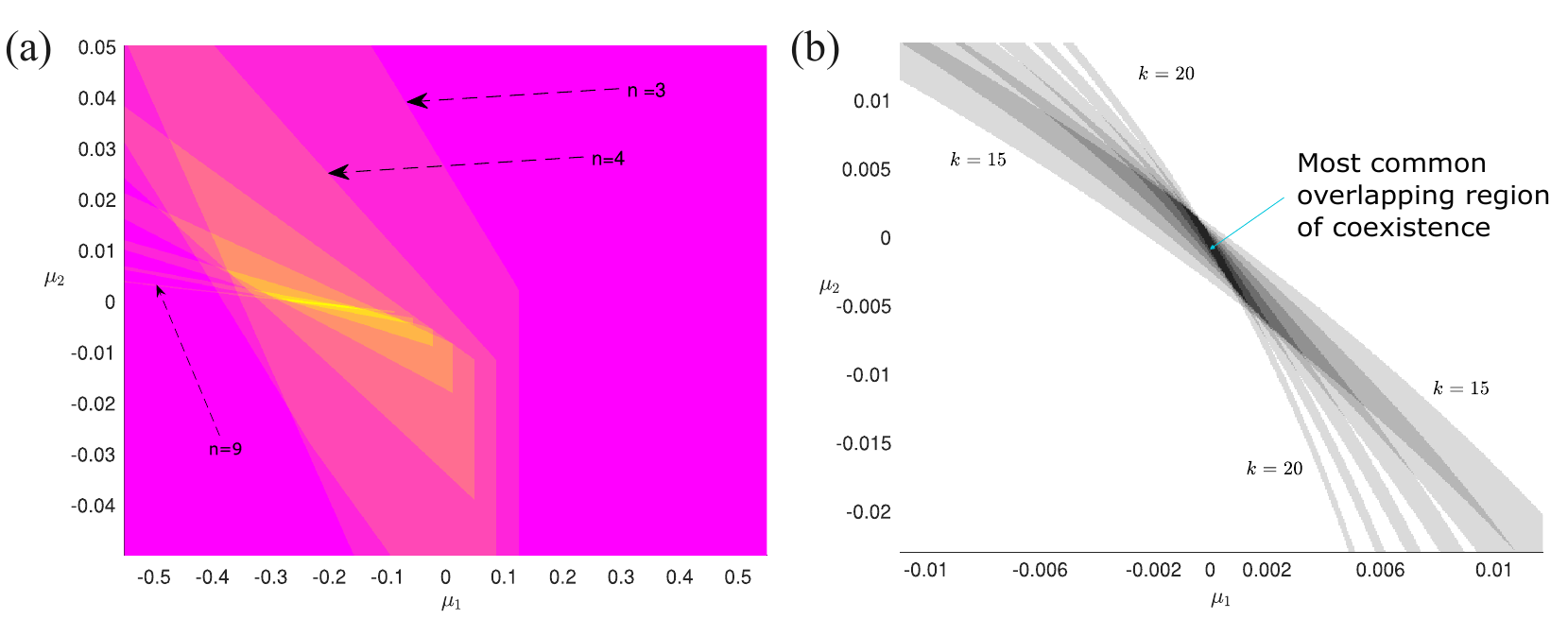}
\caption[Overlapping of stability regions from period-$16$ to period-$21$ solution in the $\mu_{1}-\mu_{2}$ parameter plane.]{The \( \mu_1 \)-\( \mu_2 \) parameter plane illustrates the overlapping stability regions for period-\((k+1)\) solutions, where \( k \) ranges from 15 to 20. The parameters are set as \( a_{1} = b_{1} = 0 \), \( \alpha = 0.8 \), \( c_{2} = -0.5 \), \( d_{5} = 1 \), \( \mu_3 = 0 \), and \( \mu_4 = 0 \).}
\label{fig:mu1mu2Overlapping}
\end{center}
\end{figure}

Figure~\ref{fig:mu1mu2Overlapping} presents the stability regions for period-$(k+1)$ solutions, considering values of \( k \) from 15 to 20. Since these regions overlap, the black-shaded area represents the domain where all six periodic solutions coexist and remain stable. In Figure~\ref{fig:mu1mu2Overlapping}(a), the overlapping region is marked with a pink tinge for periodic solutions ranging from \( n=3 \) to \( n=9 \), revealing that the most frequent overlapping area appears in yellow. \textcolor{black}{The region where several of these stability domains intersect is highlighted in yellow, indicating a zone of maximal coexistence, where multiple periodic orbits are simultaneously stable. This suggests that the system exhibits a high degree of multistability in this parameter regime.} Figure~\ref{fig:mu1mu2Overlapping}(b) highlights the predominant overlapping region for \( k=15 \) to \( k=20 \), represented in black (darker shading indicating the extent of coexistence).  \textcolor{black}{The darkest region, shaded in black, marks the most common overlapping region of coexistence, where all six periodic solutions are stable. This zone of dense overlap reflects a robust multistable structure, revealing the system’s sensitivity to small parameter changes and its potential for rich dynamic transitions.}
  
\begin{figure}[t!]
\begin{center}
\includegraphics[width=1\textwidth]{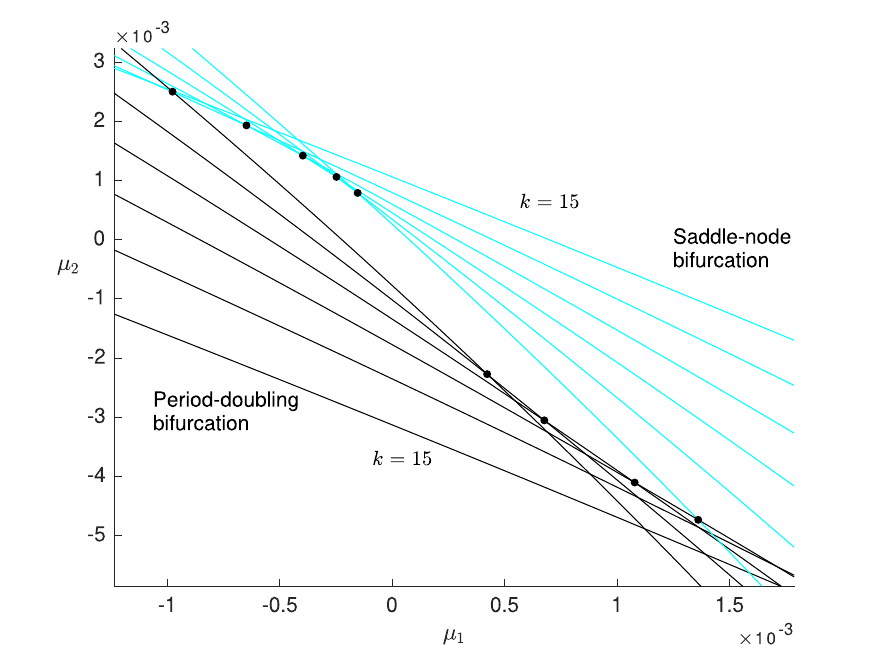}
\caption[Stability region of the periodic solutions from period $16$ to $21$ (nine-sided polygon).]{The stability regions for period-\((k+1)\) solutions, with \( k \) ranging from 15 to 20, are shown. Cyan curves represent saddle-node bifurcations, while black curves indicate period-doubling bifurcations. The parameters are set to \( \alpha = 0.8 \), \( a_{1} = b_{1} = 0 \), \( \mu_{3} = \mu_{4} = 0 \), \( c_{2} = -0.5 \), and \( d_{5} = 1 \).}
\label{fig:ninesided}
\end{center}
\end{figure}

\begin{figure}[t!]
\begin{center}
\includegraphics[width=1\textwidth]{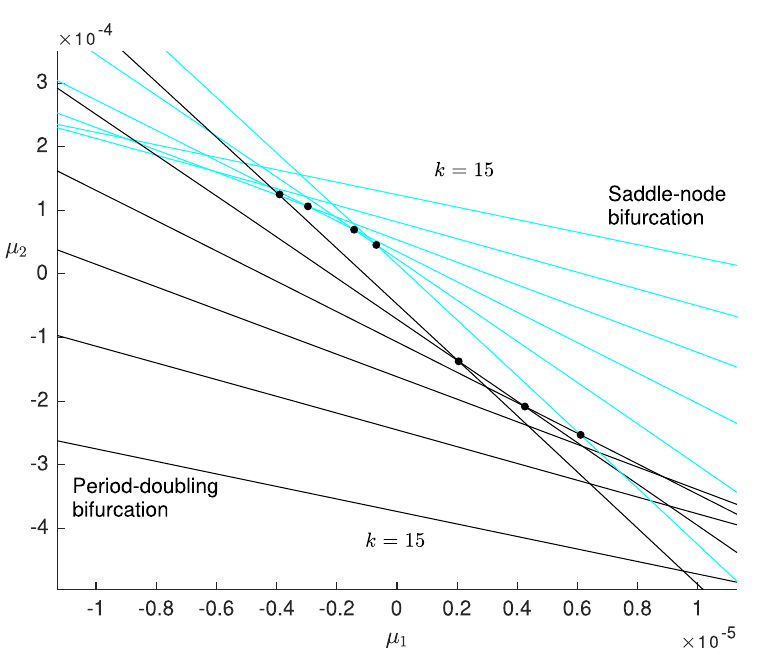}
\caption[Stability region of the periodic solutions from period $16$ to $21$ (seven-sided polygon).]{This figure shows most common overlapping region as in Fig.~\ref{fig:ninesided} but with $\alpha = 0.7$.}
\label{fig:Sevensided}
\end{center}
\end{figure}

The intricate structure of the overlapping black region is illustrated below. The saddle-node bifurcation curve (cyan) and the period-doubling bifurcation curve (black) have been computed for values of \( k \) ranging from 15 to 20 by using the analytical expressions of saddle-node ad period-doubling bifurcations, see \eqref{eq:SNeqn} and \eqref{eq:PDeqn}. For instance, in Fig.~\ref{fig:ninesided}, when \( \alpha = 0.8 \), the overlapping region forms a nine-sided polygon, with its vertices marked by black dots. In contrast, for \( \alpha = 0.7 \), this region takes the shape of a seven-sided polygon, as shown in Fig.~\ref{fig:Sevensided}. These observations highlight that regions supporting multiple stable periodic solutions often exhibit complex geometries that change with parameter variations. This motivates further exploration of how the structure of common overlapping regions evolves as parameters vary, a topic that will be examined in detail in \S \ref{sec:complexity}.

\subsection{Computation of SN and PD bifurcation points}

We have demonstrated the existence of sequences of saddle-node and period-doubling bifurcations in the vicinity of a globally resonant homoclinic tangency. Here, we numerically compute these bifurcations in the map \eqref{eq:fEx}. While numerical continuation software such as {\sc auto} \cite{DoCh07} or {\sc matcontm} \cite{KuMe19} could be used to track periodic solutions, we found that a more direct approach—akin to a manual bisection method—was sufficient for our analysis. This method involves carefully examining changes in the phase portrait, which, while straightforward, has its limitations. For a more in-depth bifurcation analysis, numerical continuation software would likely be necessary.  

\begin{figure}[t!]
\begin{center}
\includegraphics[width=0.8\textwidth]{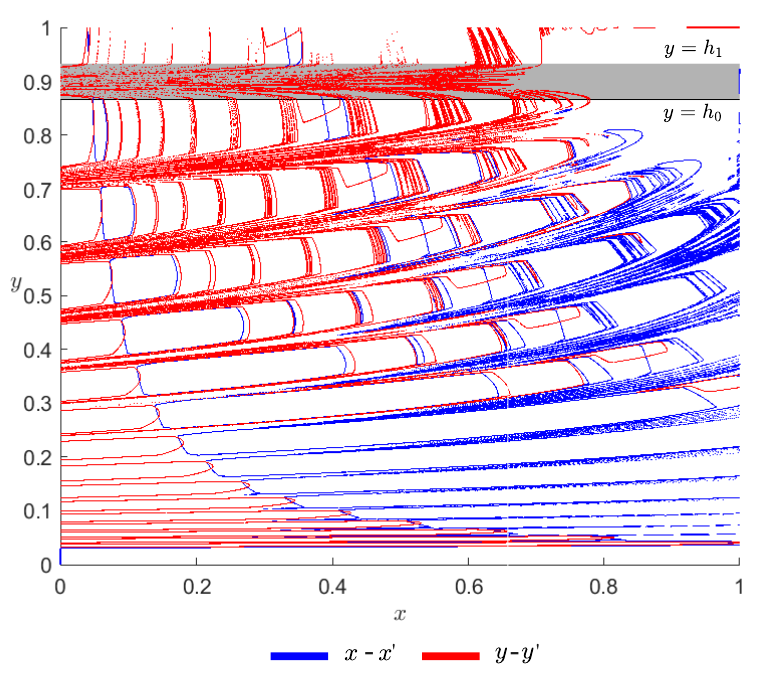}
\caption[Contour plot of the $(k+1)$th iterate of the map $f$]{Contour plot of \( f^{k+1}(x,y) = (x,y) \) for \( k=15 \), where the first component is represented in blue and the second component in red. The parameters are chosen as \( \mu_{1} = \mu_{2} = \mu_{3} = \mu_{4} = 0 \) and follow the values specified in \eqref{eq:toyParameters}.}
\label{fig:ContourPlot}
\end{center}
\end{figure}

To identify saddle-node bifurcations, we analyze contour plots of the equation \( f^{k+1}(x,y) = (x,y) \). As an example, for \( k=15 \), corresponding to a period-16 solution, we define \( (x',y') = f^{k+1}(x,y) \). In Fig.~\ref{fig:ContourPlot}, the blue curves represent locations where \( x' - x = 0 \), while the red curves indicate where \( y' - y = 0 \). These contours were generated in {\sc matlab} using the `contour' function, which evaluates \( f^{k+1}(x,y) - (x,y) \) over a fine grid and fits curves to the points where each component is zero. This method is analogous to determining nullclines in continuous dynamical systems. 

\begin{figure}[htbp!]
\begin{center}
\setlength{\unitlength}{1cm}
\vspace*{-1.6cm}\hspace*{-7cm}
\begin{picture}(5.1,22)
\put(0,-1){\includegraphics[width=10cm]{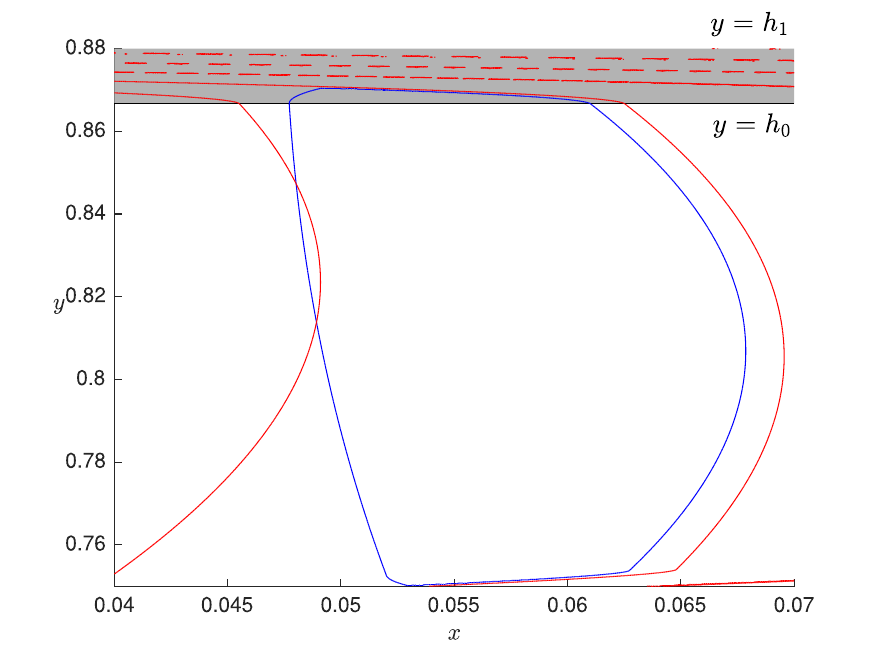}}
\put(0,6.8){\includegraphics[width=10cm]{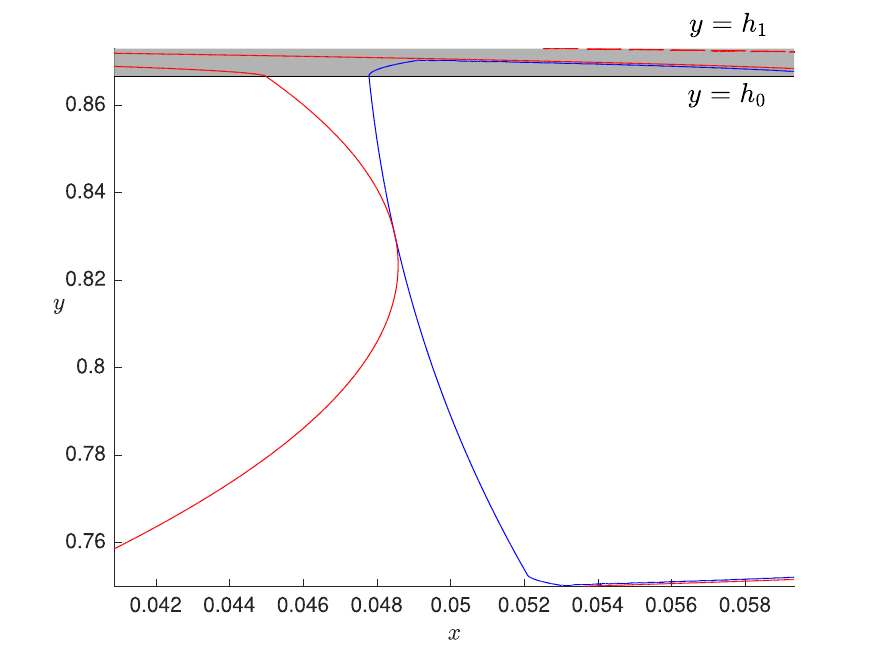}}
\put(0,14.4){\includegraphics[width=10cm]{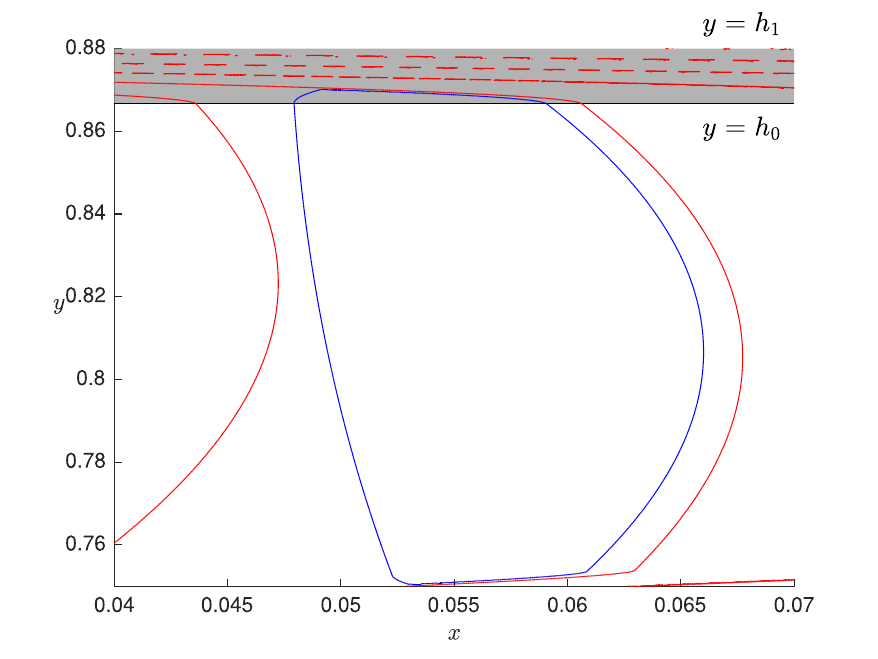}}
\put(0,6.8){\small \bf c)}
\put(0,14.4){\small \bf b)}
\put(0,21.4){\small \bf a)}
\end{picture}
\vspace*{1cm}\caption[Computation of saddle-node bifurcation numerically.]{Contour plot of \( f^{k+1}(x,y) = (x,y) \) for \( k=15 \), where the first component is shown in blue and the second in red. In (a), the contours intersect at \( \mu_{1} = 0.0005 \), indicating the presence of two fixed points. In (b), a tangential intersection occurs at \( \mu_{1} = 0.0009257 \), signifying a saddle-node bifurcation. In (c), no intersections are observed at \( \mu_{1} = 0.002 \). The parameters are chosen as \( \mu_{2} = \mu_{3} = \mu_{4} = 0 \) and follow the values specified in \eqref{eq:toyParameters}.}
	\label{fig:SNUnfold}
\end{center}
\end{figure}

Fixed points of \( f^{k+1} \), which correspond to points where the blue and red curves intersect, indicate period-\((k+1)\) solutions. Fig.~\ref{fig:SNUnfold} illustrates how the contours evolve as the parameter \( \mu_1 \) is varied, revealing the occurrence of saddle-node bifurcations. Specifically, at \( \mu_{1} = 0.0005 \), two fixed points are present, which gradually move closer together as \( \mu_{1} \) increases. At approximately \( \mu_{1} = 0.0009257 \), these points collide, signaling a saddle-node bifurcation. This critical point is highlighted in Fig.~\ref{fig:SNUnfold} (b). Additional saddle-node bifurcation points are determined using the same procedure.

To determine period-doubling bifurcations, we begin with an initial estimate near the periodic solution and track how the forward orbit behaves as parameters are varied. If the orbit converges to the periodic solution, we then compute the eigenvalues of the Jacobian matrix for the map \( f^{k+1} \). As a period-doubling bifurcation is approached, one of these eigenvalues tends toward \(-1\).  

\begin{figure}[htbp!]
\begin{center}
\setlength{\unitlength}{1cm}
\hspace*{-3cm}
\begin{picture}(12,5.1)
\put(0,0){\includegraphics[width=0.37\textwidth]{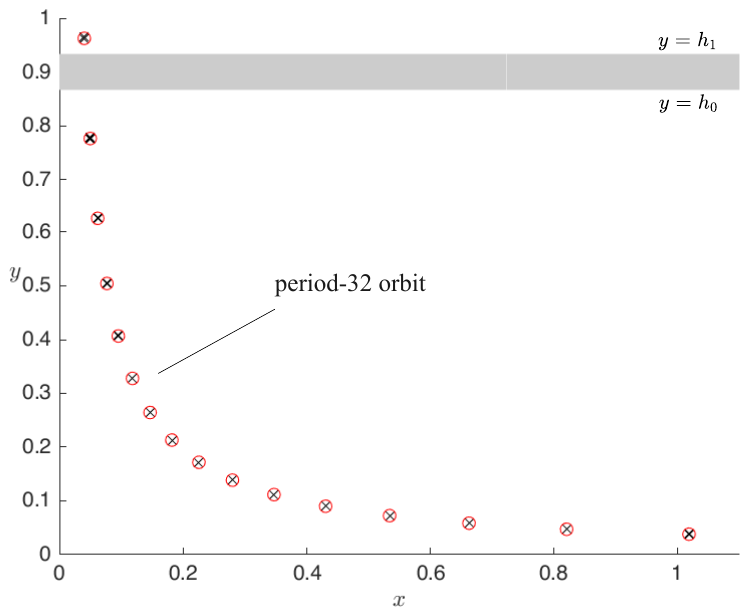}}
\put(8.6,0){\includegraphics[width=0.37\textwidth]{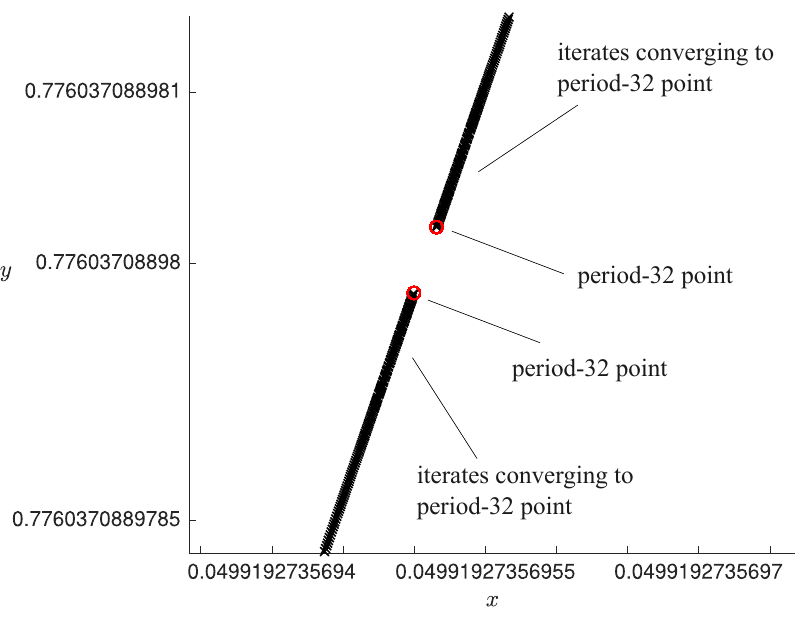}}
\put(-0.4,5.1){\small \bf a)}
\put(7.6,5.1){\small \bf b)}
\end{picture}
\caption[Computation of period doubling bifurcation numerically.]{ In (a), the orbit is represented by red circles, while the iterates are marked with black crosses, showing convergence to a stable period-32 orbit. In (b), a magnified view near a periodic point reveals two closely spaced points, indicating that a period-16 orbit has recently undergone a period-doubling bifurcation. The red circles highlight the final iterates of the computed orbit. In panel (b), the black crosses appear as straight lines due to their proximity but actually represent a discrete set of points converging toward the two periodic points approximated by the red circles. The parameters are set as \( \mu_{2} = \mu_{3} = \mu_{4} = 0 \) and follow the values specified in \eqref{eq:toyParameters}.}
	\label{fig:PDUnfold}
\end{center}
\end{figure}

For instance, considering \( k=15 \) and varying the parameter \( \mu_1 \), Fig.~\ref{fig:PDUnfold} illustrates that at \( \mu_{1} = -0.003594 \), a supercritical period-doubling bifurcation has just occurred, leading to a doubling of the period from 16 to 32. Since the doubled orbit is not easily distinguishable, a magnified view near one of the periodic points is provided in Fig.~\ref{fig:PDUnfold} (b) for clarity.

\section{Complexity of coexistence regions of stable periodic orbits}
\label{sec:complexity}
In the previous sections, we derived analytically the expressions for saddle-node and period-doubling bifurcations by trace and determinant computations. As simpler cases, we considered $\mu_{3} = 0$ and analysed the bifurcation curves on the $\mu_{1} - \mu_{2}$ plane for various values of $k \in [k_{min},k_{max}]$ for the coexistence of $k_{max} - k_{min} + 1$ number of coexisting stable periodic orbits. However, discussions regarding the shape and sizes of the sets depicting coexistence of stable periodic orbits has not yet been studied in the literature. Near the origin, the saddle-node bifurcation and period-doubling bifurcation curves behave as straight lines. When $k$ is varied, the slope of the bifurcation curves changes in such a way that they develop a common overlapping region that denotes the coexistence of stable periodic orbits from $k_{min} \leq k \leq k_{max}$. Our aim is to understand the overlapping region of coexistence. In this study, our objective is to answer the following questions:
\begin{itemize}
\item How to identify numerically the polygon representing the most common overlapping region.
\item How the most common coexistence regions varies with dynamical parameters of the map. This is important as previous studies suggest that the shape of the co-existence region varies with parameters.
\item Analyse how the number of vertices of the polygon varies with the variation of parameters.
\item Visualize the region of coexistence on the $\mu_{1} - \mu_{2} - \mu_{3}$ axis.
\item Segmenting the most common overlapping region via the convex hull arguments and computing the area of the most common overlapping region.
\item How the area of the most common overlapping region vary with the variation of parameters.
\end{itemize}

\textcolor{black}{For parameters near the origin, the saddle-node and period-doubling bifurcation curves behave like straight lines. This has been shown in Theorem \ref{thm:snpd}. Furthermore, we show that for various periodic orbits, the most common overlapping coexistence region is a convex set and polygon constituted by straight line segments. This has been proved in Theorem \ref{th:convex} and Theorem \ref{th:polygon}}.

\begin{theorem}
For small values of $\mu_{2}, \mu_{3}$ near the origin, the saddle-node bifurcation surfaces $\mu_{1,SN}^{k} (\mu_2 ,\mu_3)$ and period-doubling bifurcation surfaces $\mu_{1,PD}^{k} (\mu_2 ,\mu_3)$, see \eqref{eq:SNeqn}, \eqref{eq:PDeqn}, are given by  
\begin{align*}
\mu_{1,SN}^{k} (\mu_2 ,\mu_3) &\approx A_{SN} + B_{k} \mu_2 + C \mu_{3}, \\
\mu_{1,PD}^{k} (\mu_2 ,\mu_3) &\approx A_{PD} + B_{k} \mu_2 + C \mu_{3},
\end{align*}
where 
\begin{align*}
A_{SN} &= -\frac{(c_{2} - 1)^2}{4d_{5}}, & A_{PD} &= -\frac{(c_{2} + 1)^2}{4d_{5}}, \\
B_{k} &= -\frac{2c_{2}k}{4d_{5}\alpha}, & C &= -\frac{2c_{2}}{4d_{5}}.
\end{align*}
Thus, $(\mu_1,\mu_2)$ in a close neighborhood of the origin behave approximately as straight planes in the $\mu_1-\mu_2-\mu_3$ space.
\label{thm:snpd}
\end{theorem}

\begin{proof}
The parameters in the bifurcation equations can be approximated for small $\mu_2, \mu_3$ and for large values of $k$. The parameter $b$ is given by  
\[ b = ((\alpha + \mu_2)c)^k. \]
Substituting $c = \frac{1}{\alpha}$, we get  
\[ b = \left(\frac{\alpha + \mu_2}{\alpha}\right)^k. \]
For small $\mu_2$, using the first-order Taylor expansion, we obtain 
\[ b \approx 1 + k\frac{\mu_2}{\alpha}. \]
Since $(\mu_3 + 1)b \approx (1 + \mu_3) (1 + k\frac{\mu_2}{\alpha})$, we expand and take the dominant terms: 
\[ (\mu_3 + 1)b \approx 1 + k\frac{\mu_2}{\alpha} + \mu_3. \]
Considering the parameter $a = c_2 b$, we obtain 
\[ a \approx c_2 (1 + k\frac{\mu_2}{\alpha}). \]
Expanding the squared terms in the numerator using a first-order approximation, we obtain $(a - C)^2 \approx a^2 - 2aC$. Substituting these approximations into \eqref{eq:SNeqn} and \eqref{eq:PDeqn}, we get 
\begin{align*}
\mu_{1,SN}^{k} &\approx -\frac{(c_2 - 1)^2}{4d_5} - \frac{2c_2k}{4d_5\alpha} \mu_2 - \frac{2c_2}{4d_5} \mu_3, \\
\mu_{1,PD}^{k} &\approx -\frac{(c_2 + 1)^2}{4d_5} - \frac{2c_2k}{4d_5\alpha} \mu_2 - \frac{2c_2}{4d_5} \mu_3.
\end{align*}
These are linear equations in $\mu_2, \mu_3$, confirming that the bifurcation curves behave as straight planes in a small neighborhood of the origin.
\end{proof}

\begin{theorem}
Let $\mu_{1,SN}^{k}(\mu_2), \mu_{1,PD}^{k}(\mu_2)$ be linear functions defined for discrete values of $k$ in the range $k_{\min} \leq k \leq k_{\max}$. Define the regions:
\begin{align*}
R_1 &= \{(\mu_2, \mu_1) \mid \mu_1 > \max_{k_{\min} \leq k \leq k_{\max}} \mu_{1,SN}^{k}(\mu_2) \}, \\
R_2 &= \{(\mu_2, \mu_1) \mid \mu_1 < \min_{k_{\min} \leq k \leq k_{\max}} \mu_{1,PD}^{k}(\mu_2) \}.
\end{align*}
Then the region of coexistence $R_1 \cap R_2$ is a convex set.
\label{th:convex}
\end{theorem}

\begin{proof}
Each $\mu_{1,SN}^{k}(\mu_2)$ is a linear function of $\mu_2$, so the upper envelope
\[
f_{SN}(\mu_2) = \max_{k_{\min} \leq k \leq k_{\max}} \mu_{1,SN}^{k}(\mu_2)
\]
is the pointwise maximum of finitely many linear functions, which is a convex function \cite{rockafellar-1970a}. Similarly, the function
\[
f_{PD}(\mu_2) = \min_{k_{\min} \leq k \leq k_{\max}} \mu_{1,PD}^{k}(\mu_2)
\]
is the pointwise minimum of finitely many linear functions, which is also convex \cite{Boyd2004}. Then, the set
\[
R_1 \cap R_2 = \{ (\mu_2, \mu_1) \in \mathbb{R}^2 \mid f_{SN}(\mu_2) < \mu_1 < f_{PD}(\mu_2) \}
\]
is the set of points lying between two convex curves. Since the sublevel and superlevel sets of convex functions are convex, the vertical strip between them is a convex set. Hence, $R_1 \cap R_2$ is convex.
\end{proof}

\begin{theorem}
The common overlapping region of coexistence $R_{1} \cap R_{2}$ is a polygon.
\label{th:polygon}
\end{theorem}

\begin{proof}
According to Theorem~\ref{thm:snpd}, for sufficiently small values of the parameters $\mu_2$ and $\mu_3$, and for a fixed integer $k$, the saddle-node and period-doubling bifurcation curves, denoted respectively by $\mu_{1,SN}^{k}(\mu_2, \mu_3)$ and $\mu_{1,PD}^{k}(\mu_2, \mu_3)$, vary affinely with respect to $\mu_2$ and $\mu_3$. As a result, when projected onto the $(\mu_2,\mu_3)$-plane, these bifurcation curves appear as planar surfaces. For a fixed value of $\mu_3$, these surfaces intersect a horizontal slice at $\mu_3$ as constant value, in straight lines within the $(\mu_1,\mu_2)$-plane.

\vspace{2mm}
Now, define two envelope functions over $\mu_2$, constructed by taking the maximum and minimum over a finite set of indices $k$:
\begin{equation}
f_{\mathrm{SN}}(\mu_2) = \max_{k_{\min} \leq k \leq k_{\max}} \mu_{1,SN}^{k}(\mu_2), \quad
f_{\mathrm{PD}}(\mu_2) = \min_{k_{\min} \leq k \leq k_{\max}} \mu_{1,PD}^{k}(\mu_2).
\end{equation}

Each of these envelope functions is piecewise linear, consisting of a finite number of linear segments, since only a finite range of $k$ values is considered and each $\mu_1$-bifurcation curve is affine in $\mu_2$.

\vspace{2mm}
The coexistence region of interest is characterized by the inequality
\begin{equation}
R_1 \cap R_2 = \left\{ (\mu_2, \mu_1) \in \mathbb{R}^2 : f_{\mathrm{SN}}(\mu_2) < \mu_1 < f_{\mathrm{PD}}(\mu_2) \right\}.
\end{equation}

This region lies vertically between the graphs of $f_{\mathrm{SN}}$ and $f_{\mathrm{PD}}$, forming a band in the $(\mu_2, \mu_1)$-plane. Since both bounding functions are made up of finitely many straight line segments, the upper and lower boundaries of this strip are piecewise-linear. It follows that the intersection $R_1 \cap R_2$ is enclosed between two such curves, and thus the boundary of this region is itself composed of a finite collection of line segments. Consequently, $R_1 \cap R_2$ is a polygonal region determined by the bifurcation branches across the finite index range.
\end{proof}

\textcolor{black}{To understand the most common overlapping region, which is a polygon, we intend to compute the vertices of the polygon region formed by the intersection of the bifurcation curves, see Fig. \ref{fig:ninesided}  and Fig. \ref{fig:Sevensided}.  Note that there are three possible types of intersections of the bifurcation curves namely saddle-node -- saddle-node intersections ($SN_{k_{1}}-SN_{k_{2}}$), period-doubling-period-doubling intersections ($PD_{k_{1}}-PD_{k_{2}}$), saddle-node--period-doubling intersections ($SN_{k_{1}}-PD_{k_{2}}$), for $k_{1},k_{2} \in [k_{min}, k_{max}]$. We are interested in computing the number of vertices of the polygon formed by the intersection of the saddle-node bifurcation curves and the period-doubling curves, saddle-node--saddle-node bifurcation curves, and period doubling- period doubling bifurcation curves from $k = k_{min}$ to $k = k_{max}$. The next section defines the boundary points/ vertices of the most common overlapping region of the coexistence and discusses a method to automatically filter out the boundary points of the most common overlapping region.}

\subsection{Identification of coexistence region}

These selected points define the vertices of the polygon representing the border of the common overlapping region. One can observe from Fig. \ref{fig:ninesided} and Fig. \ref{fig:Sevensided} that there can be many such intersection points formed by the saddle-node and period-doubling bifurcation curves. However, we only need the intersection points that form the vertices of the polygon representing the common overlapping region via root finding methods like bisection method. Out of all the possible intersection points we filter out the boundary points or vertices of the common overlapping region as follows:
If the intersection point is below all the saddle-node bifurcation curves from $k = k_{min}$ to $k=k_{max}$ and above all the period-doubling bifurcation curves from $k = k_{min}$ to $k=k_{max}$, then the intersection point is one of the vertices of the common overlapping region.\\ 
\textbf{Definition} 
(Boundary Points of the Common Overlapping Region)  

Let $( \mu_{1,SN}^{k}(\mu_2) )$ and $( \mu_{1,PD}^{k}(\mu_2) )$ be the saddle-node and period-doubling bifurcation curves for discrete values of $ k $ in the range $( k_{\min} \leq k \leq k_{\max} )$. The common overlapping region is defined by the intersection of the regions:  
$$
R_1 = \{ (\mu_1, \mu_2) \mid \mu_1 > \max_{k_{\min} \leq k \leq k_{\max}} \mu_{1,SN}^{k}(\mu_2) \}
$$
$$
R_2 = \{ (\mu_1, \mu_2) \mid \mu_1 < \min_{k_{\min} \leq k \leq k_{\max}} \mu_{1,PD}^{k}(\mu_2) \}
$$
The \textit{boundary points} or \textit{vertices} of the polygon representing the common overlapping region are the intersection points of bifurcation curves that satisfy the following conditions:  

1. Intersection Between Saddle-Node and Period-Doubling Bifurcation Curves:\\  
   An intersection point \( (\mu_1^*, \mu_2^*) \) is a solution to  
   \[
   \mu_1^* = \mu_{1,SN}^{k}(\mu_2^*) = \mu_{1,PD}^{k'}(\mu_2^*)
   \]
   for some \( k, k' \) in the range \( [k_{\min}, k_{\max}] \), meaning it satisfies  
   \[
   \mu_{1,SN}^{k}(\mu_2) = \mu_{1,PD}^{k'}(\mu_2).
   \]

2. Intersection Between Saddle-Node Bifurcation Curves:\\  
   An intersection point \( (\mu_1^*, \mu_2^*) \) is also a boundary point if it satisfies  
   \[
   \mu_1^* = \mu_{1,SN}^{k}(\mu_2^*) = \mu_{1,SN}^{k'}(\mu_2^*)
   \]
   for some \( k \neq k' \) in the range \( [k_{\min}, k_{\max}] \), meaning it satisfies  
   \[
   \mu_{1,SN}^{k}(\mu_2) = \mu_{1,SN}^{k'}(\mu_2).
   \]

3. Intersection Between Period-Doubling Bifurcation Curves:\\
   Similarly, an intersection point \( (\mu_1^*, \mu_2^*) \) is a boundary point if it satisfies  
   \[
   \mu_1^* = \mu_{1,PD}^{k}(\mu_2^*) = \mu_{1,PD}^{k'}(\mu_2^*)
   \]
   for some \( k \neq k' \) in the range \( [k_{\min}, k_{\max}] \), meaning it satisfies  
   \[
   \mu_{1,PD}^{k}(\mu_2) = \mu_{1,PD}^{k'}(\mu_2).
   \]

4. Filtering the True Boundary Points  
   From all possible intersection points, we select only those that satisfy one of the following: \\ 
   - The point is below all the saddle-node bifurcation curves, i.e.,  
     \[
     \mu_1^* \leq \mu_{1,SN}^{k}(\mu_2^*) \quad \forall k \in [k_{\min}, k_{\max}].
     \]
   - The point is \textit{above all} the period-doubling bifurcation curves, i.e.,  
     \[
     \mu_1^* \geq \mu_{1,PD}^{k}(\mu_2^*) \quad \forall k \in [k_{\min}, k_{\max}].
     \]
and such points refer to as \textit{boundary points} of the most common overlapping region.

After understanding the concept of boundary points or vertices of the most common overlapping region, we describe an algorithm to detect the vertices / boundary points of the most common overlapping region, see Algorithm \ref{alg:vertices}. 

\begin{algorithm}[htbp]
\caption{Detection of the vertices of the overlapping region}
\SetKwInOut{Input}{Input}
\SetKwInOut{Output}{Output}
\SetKwComment{Comment}{\# }{}

\Input{Saddle-node and period-doubling bifurcation curves, parameter range \( \mu_2 \), bifurcation index \( k \) values}
\Output{Intersection points categorized as: 
\begin{itemize}
  \item Saddle-node $\cap$ Saddle-node intersections
  \item Period-doubling $\cap$ Period-doubling intersections
  \item Saddle-node $\cap$ Period-doubling intersections
  \item Filtered vertices of the enclosed region
\end{itemize}}

\BlankLine

\textbf{STEP 1:} Generate bifurcation curves\;
Generate \( \mu_2 \) over the given range\;
\ForEach{\( k \) in the bifurcation index range}{
    Compute \( \mu_1 \) for each \( \mu_2 \) using:\;
    \Indp
    Saddle-node bifurcation: \( f_{\text{SN}}(\mu_2, k) \)\;
    Period-doubling bifurcation: \( f_{\text{PD}}(\mu_2, k) \)\;
    \Indm
}

\BlankLine

\textbf{STEP 2:} Find intersections of bifurcation curves\;

\textbf{(a) Saddle-node $\cap$ Saddle-node:}\;
\ForEach{pair \( (k_1, k_2) \) with \( k_1 \neq k_2 \)}{
    Detect sign changes in \( f_{\text{SN}}(\mu_2, k_1) - f_{\text{SN}}(\mu_2, k_2) \)\;
    Refine intersection points using \texttt{fsolve}\;
    Store valid intersection points\;
}

\textbf{(b) Period-doubling $\cap$ Period-doubling:}\;
\ForEach{pair \( (k_1, k_2) \) with \( k_1 \neq k_2 \)}{
    Detect sign changes in \( f_{\text{PD}}(\mu_2, k_1) - f_{\text{PD}}(\mu_2, k_2) \)\;
    Refine intersection points using \texttt{fsolve}\;
    Store valid intersection points\;
}

\textbf{(c) Saddle-node $\cap$ Period-doubling:}\;
\ForEach{pair \( (k_1, k_2) \)}{
    Detect sign changes in \( f_{\text{SN}}(\mu_2, k_1) - f_{\text{PD}}(\mu_2, k_2) \)\;
    Refine intersection points using \texttt{fsolve}\;
    Store valid intersection points\;
}

\BlankLine

\textbf{STEP 3:} Filter intersections to find enclosed region vertices\;
\ForEach{intersection point \( (\mu_1, \mu_2) \)}{
    Check if point lies below all saddle-node curves\;
    Check if point lies above all period-doubling curves\;
    \If{both conditions satisfied}{
        Retain the point as a vertex of the enclosed region\;
    }
}
\label{alg:vertices}
\end{algorithm}

\textcolor{black}{After computing the vertices/ boundary points of the most common overlapping region, we next discuss an algorithm which can compute the area of any sided polygon via the use of Shoelace algorithm, see Algorithm \ref{alg:area}. We would like to analyze the variation of the area of the most common overlapping region with the variation of parameters.}

\begin{algorithm}[htbp]
\caption{Computing area of the polygon with unsorted vertices}
\SetKwInOut{Input}{Input}
\SetKwInOut{Output}{Output}

\Input{A set of unordered vertices \( \{(x_1, y_1), (x_2, y_2), \dots, (x_n, y_n)\} \)}
\Output{Area of the polygon enclosed by the given vertices}

\textbf{Step 1: Compute centroid} \( (C_x, C_y) \):\\
\[
C_x = \frac{1}{n} \sum_{i=1}^{n} x_i, \quad C_y = \frac{1}{n} \sum_{i=1}^{n} y_i
\]

\textbf{Step 2: Sort vertices by angle from centroid:}\\
\For{each point \( (x_i, y_i) \)}{
    Compute \( \theta_i = \text{atan2}(y_i - C_y, x_i - C_x) \)
}
Sort the points in increasing order of \( \theta_i \)

\textbf{Step 3: Compute area using shoelace formula:}\\
\If{\( n < 3 \)}{
    Raise error: A polygon must have at least 3 vertices
}
Initialize \( \text{area} = 0 \)\\
\For{each consecutive pair \( (x_i, y_i), (x_{i+1}, y_{i+1}) \)}{
    \( \text{area} = \text{area} + (x_i y_{i+1} - x_{i+1} y_i) \)
}
Compute final area: \( A = \frac{1}{2} | \text{area} | \)

\Return \( A \)
\label{alg:area}
\end{algorithm}

\begin{algorithm}[htbp]
\caption{Finding and Connecting Disjoint Regions}
\SetKwInOut{Input}{Input}
\SetKwInOut{Output}{Output}

\Input{Set of boundary points \( \{p_1, p_2, \dots, p_n\} \)}
\Output{Connected set of points representing the boundary of the region}

\textbf{Step 1: Find closest points between regions} \\
Initialize \( \text{min\_dist} = \infty \), \( \text{closest\_pts} = \text{None} \)\\
\For{each \( p_1 \) in region1}{
    \For{each \( p_2 \) in region2}{
        Compute \( \text{dist} = \sqrt{(x_1 - x_2)^2 + (y_1 - y_2)^2} \)\\
        \If{\( \text{dist} < \text{min\_dist} \)}{
            Update \( \text{min\_dist} = \text{dist} \), \( \text{closest\_pts} = (p_1, p_2) \)
        }
    }
}

\textbf{Step 2: Connect regions} \\
Sort points by x-coordinate \\
Compute differences: \( \text{x\_diffs} = \text{diff(sorted\_points[:,0])} \)\\
Compute threshold: \( \text{gap\_threshold} = \text{mean} + 2 \times \text{std} \)\\
Identify gap indices where \( \text{x\_diffs} > \text{gap\_threshold} \)

\If{no gap indices}{
    \Return original points
}

\For{each gap index \( i \)}{
    Let \( p_1 = \text{sorted\_points}[i], p_2 = \text{sorted\_points}[i+1] \)\\
    \For{each \( t \in \text{linspace}(0, 1, 5)[1:-1] \)}{
        Compute \( \text{new\_point} = p_1 + t (p_2 - p_1) \)\\
        Add \( \text{new\_point} \) to list
    }
}
\If{new points added}{
    Combine all points: \( \text{all\_points} = \text{vstack(points, new\_points)} \)\\
    Compute convex hull: \( \text{hull} = \text{ConvexHull(all\_points)} \)\\
    Extract: \( \text{hull\_points} = \text{all\_points[hull.vertices]} \)\\
    \Return \( \text{hull\_points} \)
}

\Return original points

\label{alg:disjoint}
\end{algorithm}

\subsection{Variation of the region of coexistence with parameters}
\begin{figure}[!htbp]
    \centering
    \includegraphics[width=0.8\linewidth]{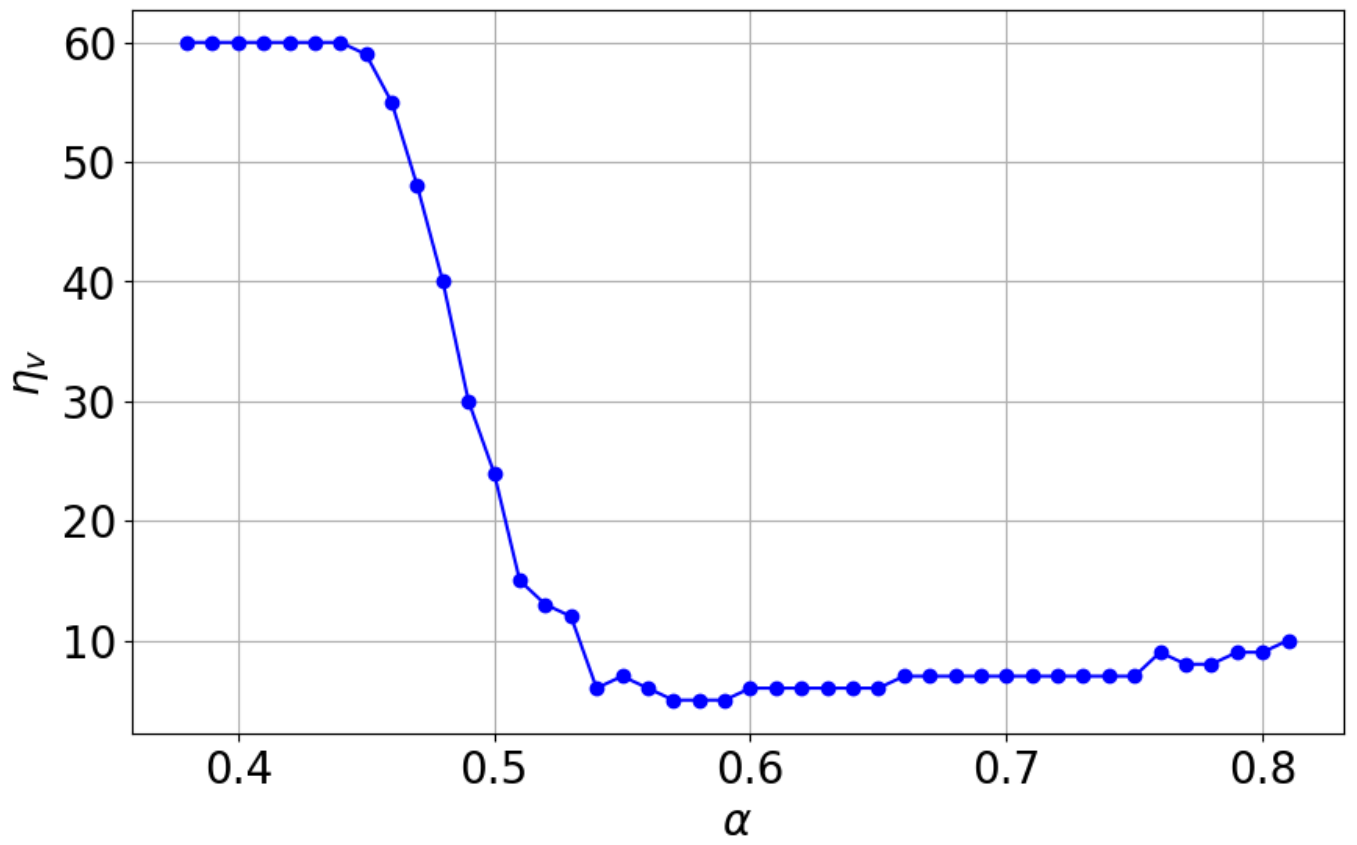}
    \caption{Variation of the number of vertices $\eta_v$ with parameter $\alpha$. With increase in $\alpha$, we observe that the number of vertices of the most common overlapping region decreases. }
     \label{fig:vertices}
\end{figure}

We observe how the number of vertices $\eta_{v}$ vary with the change in parameter $\alpha$, see Fig. \ref{fig:vertices}. We observe that for low values of $\alpha$, the number of vertices of the common overlapping region is very high. With increase in $\alpha$, the number of vertices decreases drastically. \textcolor{black}{For lower values of $\alpha$ (approximately $\alpha < 0.46$), the number of vertices remains at a constant high value of around 60, indicating a highly fragmented or complex polygonal structure in the bifurcation overlap region. This suggests that many bifurcation branches are contributing actively to the envelope functions in this range, leading to frequent switching between dominating segments. As $\alpha$ increases past a critical threshold near $\alpha \approx 0.47$, there is a rapid drop in the number of vertices, which levels off near $\eta_v \approx 6$–$10$ for $\alpha \in (0.55, 0.75)$. This sharp transition implies a simplification in the bifurcation landscape, where fewer branches are contributing and the envelope functions are dominated by only a few segments.
Interestingly, for larger values of $\alpha$ (e.g., $\alpha > 0.75$), a mild upward trend in $\eta_v$ is observed, which may suggest the re-entry or emergence of additional intersecting bifurcation curves, albeit at a less intense rate compared to the initial decrease}. A deeper dive into this observation and mathematically analysing the geometric concepts behind this seems promising future research direction.  

Over the same parameter range $\alpha$, we also analyse how the area of the common overlapping region varies. To compute the area, we employ the Shoelace algorithm \cite{Allgower1986-om}, which takes as input the set of vertices of the common overlapping region and outputs the area of the polygon formed by the common overlapping region. The idea behind the shoelace algorithm is simple. First we compute the centroid of the common overlapping region $(C_{x}, C_{y})$.  We next sort the vertices with respect to the angle it makes with the centroid (that is increasing order of the angle $\theta_{i}$). After that iteratively compute the area via the area $A = \sum_{i=1}^{n-1} (x_{i}y_{i+1} - x_{i+1}y_{i})$, where $(x_{i},y_{i})$ are the vertices of the common overlapping region. 

\begin{figure}[!htbp]
    \centering
    \includegraphics[width=0.8\linewidth]{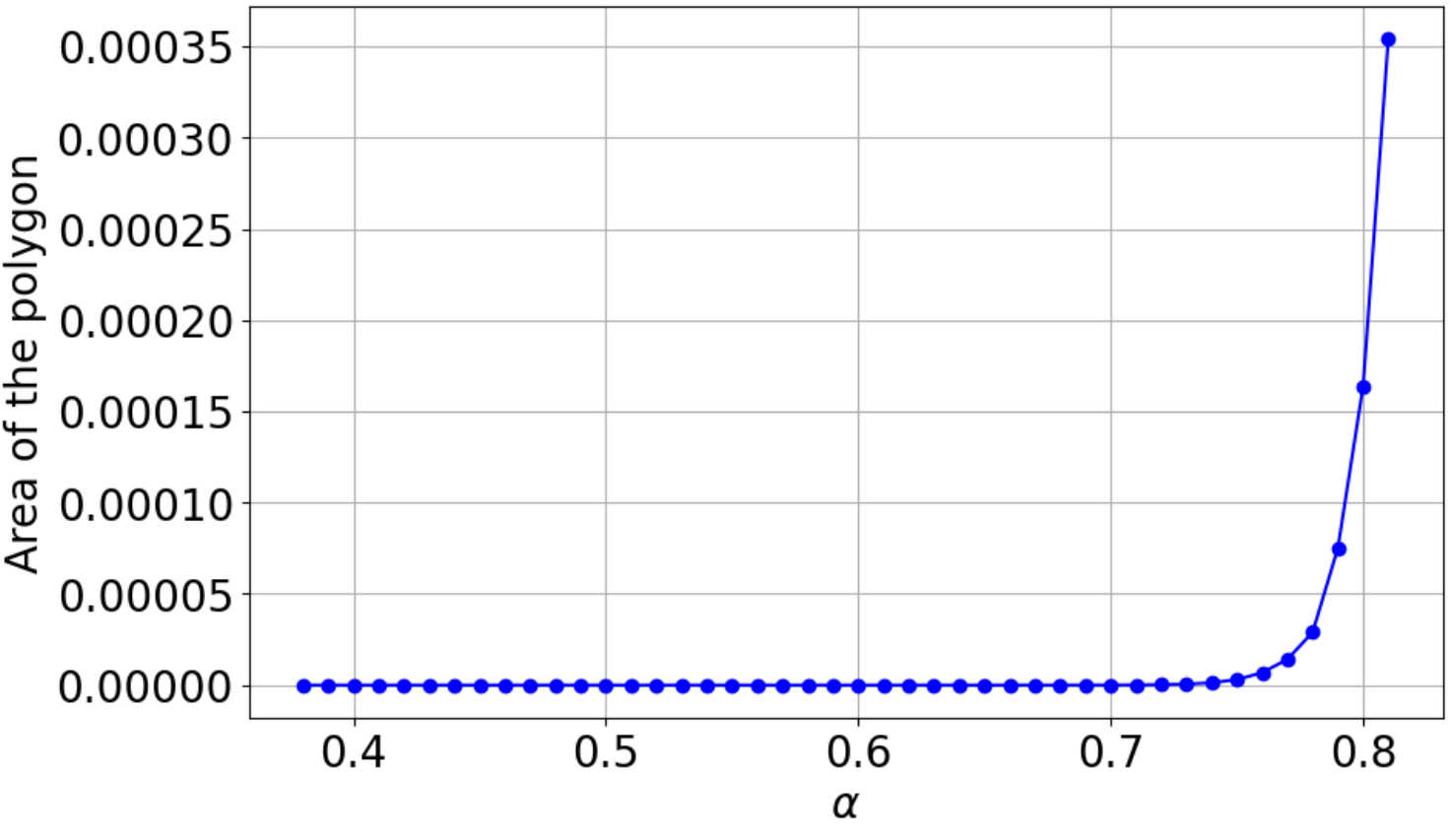}
    \caption{Variation of the area of the most common overlapping region with respect to parameter $\alpha$. We can observe that with increase in the value of $\alpha$, the area of the most common overlapping region increases. }
    \label{fig:AlphaArea}
\end{figure}

We first analyse how the area of the common overlapping region varies with parameter $\alpha$ directly related to the eigenvalues of the saddle fixed point $O$. Note that for low values of $\alpha$, the area is very close to zero (meaning the region is tiny), whereas for large values of $\alpha$, the area is bigger. We observed that for low values of $\alpha$, the area was shrinking to zero with the highest number of vertices, whereas for larger values of $\alpha$, the area was larger with less number of vertices. \textcolor{black}{Conversely, for larger values of $\alpha$, the area increases significantly, suggesting that fewer bifurcation curves dominate and intersect in a manner that encloses a much broader polygonal region, see Fig. \ref{fig:AlphaArea}. This area growth is not linear; rather, it exhibits a sharp increase after a threshold near $\alpha \approx 0.75$, hinting at a nonlinear transition in the underlying bifurcation geometry. This could be associated with a reorganization in the folding structure of the stable and unstable manifolds or the onset of more coherent resonance conditions.
This dual observation—namely, a large number of vertices but vanishing area for small $\alpha$ and fewer vertices but growing area for large $\alpha$—is particularly interesting. It suggests a transition from a highly fragmented bifurcation structure to a more coherent and spatially extensive bifurcation overlap, potentially tied to deeper dynamical transitions which remains to be uncovered.}

\begin{figure}
    \centering
    \includegraphics[width=1\linewidth]{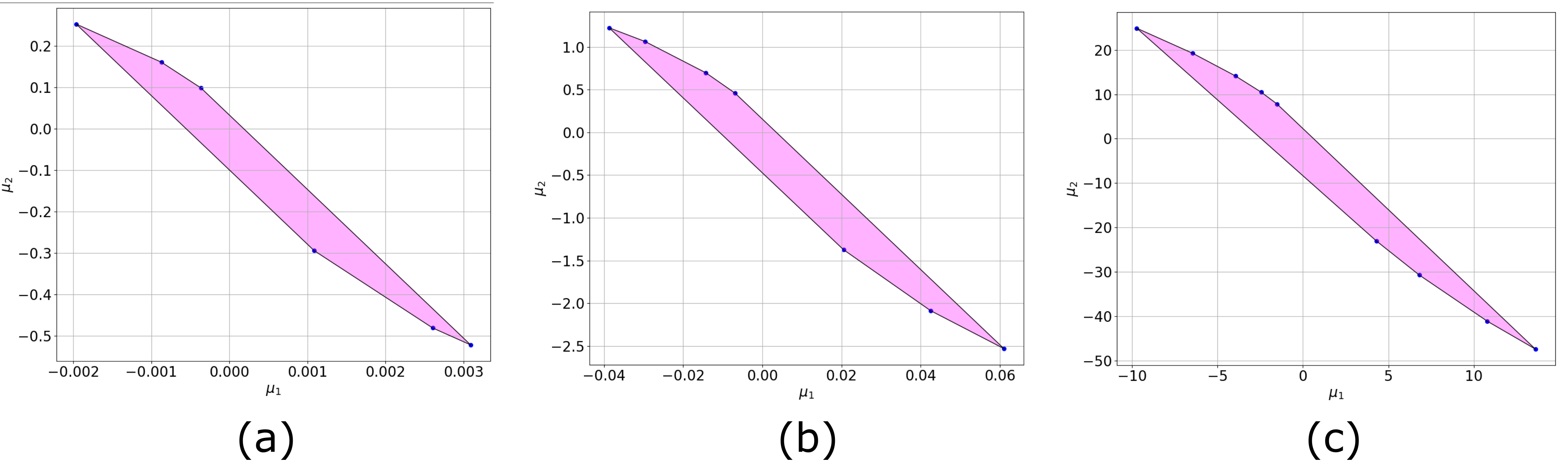}
    \caption{Illustration of the most common overlapping region of coexistence in the $\mu_1-\mu_2$ plane from $k=15$ to $k=20$ for various values of parameter $\alpha$. In (a), the most common overlapping region of coexistence is shown denoting a six sided polygon for $\alpha = 0.65$. In (b), the most common overlapping region of coexistence is shown denoting a seven sided polygon for $\alpha = 0.6$. In (c), the most common overlapping region of coexistence is shown denoting a nine sided polygon for $\alpha = 0.7$. We observe that for minor change in parameter $\alpha$, the number of vertices of the most common overlapping polygon changes drastically.}
    \label{fig:PinkShape}
\end{figure}

\textcolor{black}{To illustrate the complexity of the most common overlapping region, we showcase the most common region of coexistence from period-16 to period-21 orbit on the $\mu_1-\mu_2$ plane. For $\alpha = 0.65$, we can observe that the most common overlapping region in the $\mu_1-\mu_2$ parameter space is that of a six sided polygon, see Fig. \ref{fig:PinkShape}(a). For $\alpha = 0.6$, we observe the most common overlapping region to be a seven sided polygon, see Fig. \ref{fig:PinkShape}(b). When $\alpha$ increases to $0.7$, the most common overlapping region of coexistence is a nine sided polygon, see Fig. \ref{fig:PinkShape}(c). We can observe that it is not trivial to account for the variation of the number of vertices of the most common overlapping polygonal region of coexistence of stable periodic orbits.}  

\begin{figure}
    \centering
    \includegraphics[width=1\linewidth]{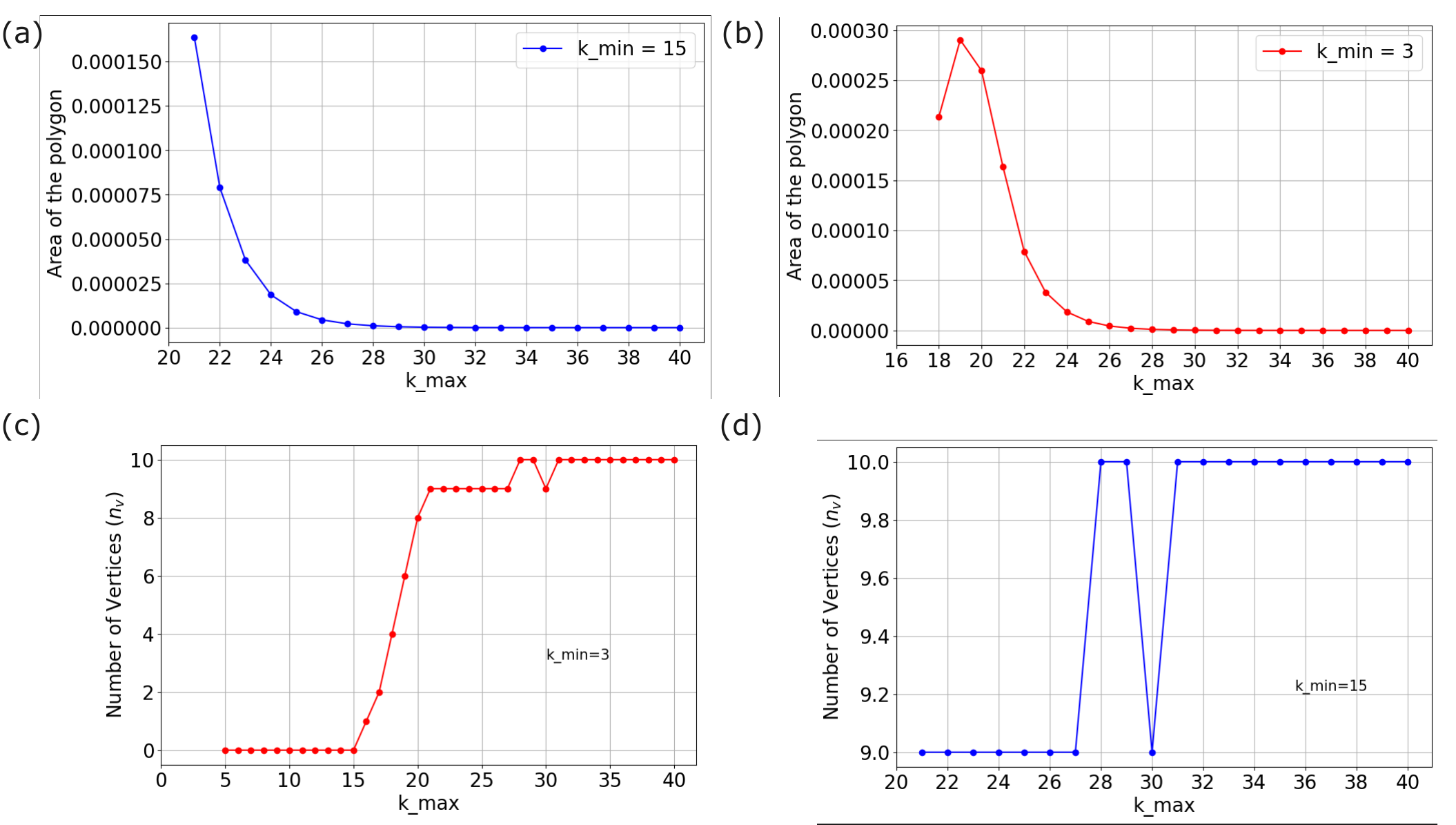}
    \caption{Illustration of the variation of area of most common overlapping region and number of vertices of the common overlapping polygon for a given number of coexisting stable periodic orbits in $[k_{min}, k_{max}]$, with $k_min = 3, 15$ as fixed and vary $k_{max}$. In (a), for fixed $k_{min} = 15$, we vary $k_{max}$ and analyse the area and number of vertices of the most common overlapping region. Observe that the area of the common overlapping region decreases with increase in $k_{max} \in [20,40]$ while the number of vertices increases with increase in $k_{max}$, see (d). In (b), when $k_{min} = 3$, we observe that the area of the polygon decreases as number of coexisting stable periodic orbits increases, see (b), while the number of vertices of the common overlapping region increases, see (c).}
    \label{fig:15}
\end{figure}

\textcolor{black}{We investigate how both the area and geometric structure (in terms of vertices) of the most frequently overlapping region change as the number of coexisting stable periodic orbits increases. This corresponds to expanding the range $[k_{\min}, k_{\max}]$. In Fig.~\ref{fig:15}(a), where $k_{\min} = 15$ is fixed and $k_{\max}$ increases from 20 to 40, there is a clear and rapid decline in the area of the common region. This trend indicates that as the number of coexisting periodic orbits grows, the region in parameter space where they all overlap becomes increasingly narrow, suggesting a potential fragmentation or a breakdown of shared structural stability across orbits.
A comparable trend can be observed in Fig.~\ref{fig:15}(b), with $k_{\min} = 3$ and $k_{\max}$ varying from 16 to 40. Although the lower bound of the period range is smaller in this case, the overlapping area still diminishes quickly, emphasizing that the presence of a large number of distinct stable orbits restricts the shared parameter space significantly.
Interestingly, the complexity of the shape of the overlapping region—as measured by the number of polygonal vertices—tells a different story, as shown in Figs.~\ref{fig:15}(c) and (d). In both scenarios, the vertex count rises with increasing $k_{\max}$, implying that while the size of the shared region contracts, its boundary becomes more intricate. This likely results from additional intersections between different periodicity regions as $k_{\max}$ increases. A marked change can be seen in Fig.~\ref{fig:15}(c) near $k_{\max} \approx 20$, where the number of vertices abruptly jumps from 2 to around 9, indicating a structural reconfiguration.
In the case shown in Fig.~\ref{fig:15}(d), with $k_{\min} = 15$, the number of vertices mostly remains at 10 across a broad range of $k_{\max}$ but briefly drops near $k_{\max} = 29$ before recovering. This abrupt change might be attributed to an atypical bifurcation event or the disappearance of a boundary segment in the overlapping region, hinting at a local structural anomaly that deserves further exploration.
These observations highlight a nuanced interplay: as the range of coexisting periodicities broadens, the shared region shrinks in size yet becomes more geometrically intricate. This suggests that the coexistence of many periodic orbits leads to a highly delicate structural condition in parameter space.
Such behavior may correspond to a rearrangement in the configuration of resonance or Arnold tongues, where the overlap zones grow increasingly thin or angular—boosting the vertex count without adding much area to the region of common stability.} 

\subsection{Two-parameter variation of the coexisting regions}
\begin{figure}
    \centering
    \includegraphics[width=0.9\linewidth]{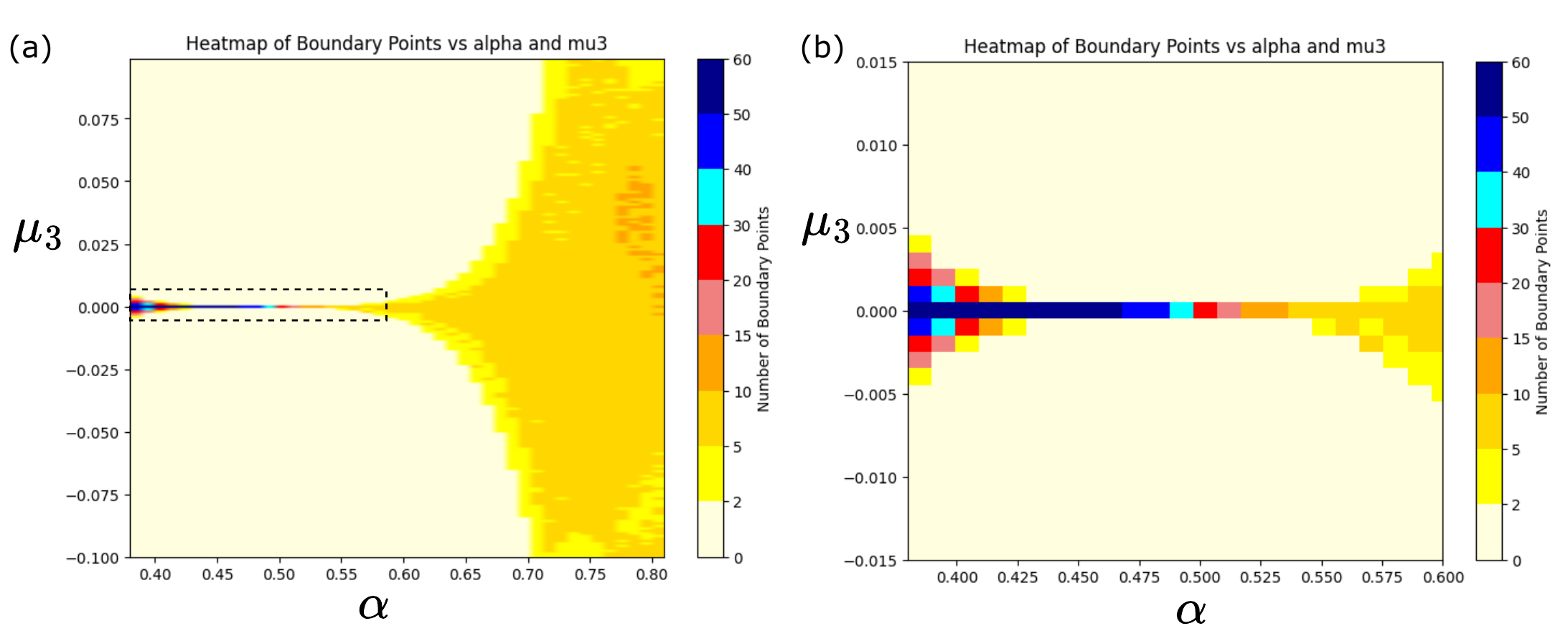}
    \caption{Variation of the vertices of the common overlapping polygon with respect to parameters $\mu_{3}$ and $\alpha$. In (a), we observe that in a tiny strip near $\mu_{3} =0$, we observe variation of colours representing the variation in the number of vertices of the common overlapping region. In (b), a zoomed version near the strip $\mu_{3} =0$ is shown. We can understand the complexity of the variation of the number of vertices of common overlapping region.}
    \label{fig:mu3alpha}
\end{figure}

\textcolor{black}{In our investigation of the shape and structure of the coexisting region of stable periodic orbits, we initially set $\mu_3 = 0$ to simplify the analysis and explore the stability region within the $(\mu_1, \mu_2, \mu_3)$-space governed by Eqs. \eqref{eq:SNeqn} and \eqref{eq:PDeqn}. However, to understand the sensitivity of the geometry of the overlapping region to variations in the third parameter $\mu_3$, we further examine how the number of vertices of the common region evolves with simultaneous changes in $\mu_3$ and $\alpha$, as shown in Fig.\ref{fig:mu3alpha}.}

In panel (a), we observe that for most values of $\mu_3$, the number of vertices remains low (mostly yellow, indicating values around 2–10), suggesting that the overlapping region retains a simple polygonal structure. However, a sharp transition in structure occurs within a very narrow strip around $\mu_3 = 0$. This strip exhibits a highly nontrivial variation in the number of vertices, reflected in the multicolored bands aligned along the $\alpha$-axis.

The zoomed-in panel (b) highlights this region of interest more clearly. Here, as $\alpha$ ranges from approximately 0.4 to 0.8, we see an alternating pattern of red, blue, and other hues corresponding to polygon vertex counts as high as 50–60. This intricate fluctuation implies that near $\mu_3 = 0$, the common region undergoes topological transitions—perhaps due to bifurcation boundary crossings or tangencies—resulting in a highly fragmented, possibly fractal-like, boundary structure.

Importantly, these high vertex counts are confined to an extremely thin strip in $\mu_3$, suggesting that while such complex overlapping regions exist, they are structurally delicate and highly sensitive to parameter perturbations. This aligns with the behavior expected in systems exhibiting infinite coexistence of stable periodic orbits, such as the GRHT map.

\section{Conclusion}
In this contribution, we revisited the planar GRHT map, which displays infinite coexistence of stable periodic orbits via the phenomenon of \textit{Globally Resonant Homoclinic Tangencies}. The main contribution of this study is to investigate the size, shape, and area of the coexisting regions of periodic orbits which was not studied in detail before in the literature. We found that the bifurcation curves behave as straightlines near the orgiin and thus near by origin, the coexisting regions take the shape of polygon with their vertices as boundary points. We illustrate that the number of vertices of the coexisting overlapping region deccreases with increase in the stable eigenvalue of the saddle point. Moreover, on contrary, the area of the coexisting overlapping region increases with increase in the stable eigenvalue of the saddle fixed point. We show that the coexisting overlapping polygonal region is a  convex set.  We discuss that it is non-trivial to compute and detect the vertices of the most common overlapping region of coexistence. To compute the number of vertices, we developed an algorithm which computes all the possible intersections of the bifurcation curves and then filters out the boundary points/ vertices of the most common overlapping region of coexistence of stable periodic orbits. After the vertices are automatically found for any set of parameters are varied, we can then develop an algorithm to compute the area of the most common overlapping region via the shoelace algorithm. Via the convex hull arguments, we develop an algorithm to automatically detect the most common overlapping region. We also understand the complexity in the shape of the most common overlapping region of coexistence with the variation of two parameters. We have varied parameters $\mu_3$ and $\alpha$ and observe that near $\mu_3 =0$, there exists a strip which showcases a colourful region which indicates complex changes in the number of vertices of the most common overlapping region of coexistence of stable periodic orbits. However, all the analysis done in the current study was for the codiemsnion-three case. It would be interesting and a bit challenging to understand the geometric properties of the coexisting regions for the codimension-four case. Moreover, we believe that the map with extreme multistability can be used for the purpose of image, audio, and video multimedia encrytpion. 

So far, only periodic orbits are explored for the GRHT map. However other dynamical behaviors need to be explored like the presence of chaotic, hyperchaotic attractors along similar lines \cite{MuniLV,MuniPersist}. Various pathways towards hyperchaos can be understood via the continuation of saddle periodic orbits \cite{MuniHC,MuniMem}. The presence of mode-locked periodic orbits would be very interesting as it can be first example of nested and coexisting of several mode-locked periodic orbits resulting to either saddle-node or saddle-focus connection \cite{MuniErgodicResonant}. This can open doors to understand bifurcation scenarios of coexisting mode-locked periodic orbits \cite{MuniHR}. Ring-star network of GRHT map can be considered \cite{MuniChua}. It would be interesting to understand how the infinite coexistence plays a role in illustration of network behavior of GRHT map. Detailed two parameter Lyapunov charts remains to be carried out in the case of GRHT map \cite{Ramrezvila2024}. This can lead to the identification of various other dynamical regimes and also can lead to an understanding of  distribution of various coexisting periodicity regions. \textcolor{black}{The electronic implementation of the GRHT mapping and experimental verification is an undergoing current work. Physical systems exhibiting such complex dynamics as exhibited by the GRHT mapping has not yet been reported and remains a future work}.

\section*{Acknowledgements}
S.S.M expresses his thanks to his M.Sc. student J. S. Ram who discussed some parts of the manuscript with me. Such discussion led to new ideas and algorithms to pursue and understand the complexity regions of coexistence. 

\section*{Credit author statement}
\textbf{Sishu Shankar Muni}: Writing, editing, software, Supervision, Conceptualization, Writing(review, editing), Visualization, Project Administration.

\section*{Conflict of Interest}
The authors confirm that this work is free from conflicts of interest.
\section*{Data Availability}

The data can be provided on reasonable request from the corresponding author.

\bibliographystyle{unsrt} 
\bibliography{Arxiv}
\end{document}